\documentclass[10pt,journal,compsoc]{IEEEtran}

\usepackage{hyperref}%
\hypersetup{colorlinks=true, linkcolor=blue, breaklinks=true, urlcolor=blue, citecolor=blue}

\usepackage{colortbl}
\definecolor{LightGray}{rgb}{0.9, 0.9, 0.9}

\usepackage{amsmath,amsfonts,amssymb}
\usepackage{cite}

\usepackage{mathtools}
\usepackage{CJK}
\usepackage{listings}

\usepackage{tikz}

\usepackage{booktabs}
\usepackage{pgfplots}
\pgfplotsset{compat=newest}
\usetikzlibrary{pgfplots.groupplots}
\usepgfplotslibrary{groupplots}
\usepgfplotslibrary{ternary}
\usepackage{pgfplotstable}
\usepackage{multirow}
\usetikzlibrary{arrows}
\usepgflibrary{arrows}
\usetikzlibrary{tikzmark,overlay-beamer-styles}
\usetikzlibrary{arrows.meta, positioning, matrix}

\usepackage{stfloats}
\usepackage{subcaption}

\usepackage{comment}
\usepackage{balance}

\title{Opinion Dynamics Models for\\ Sentiment Evolution in Weibo Blogs}
\author{Yulong He, Anton V. Proskurnikov,~\IEEEmembership{Senior~Member,~IEEE},  and Artem Sedakov
\thanks{Yulong He and Artem Sedakov are with St. Petersburg State University, St. Petersburg, Russia 199034
	{\tt\{st065924@student.spbu.ru; a.sedakov@spbu.ru}. }
\thanks{Anton V. Proskurnikov is with Politecnico di Torino, Turin, Italy, 10129	
	~\tt{anton.p.1982@ieee.org}.}
}

\begin{document}
	
	\maketitle
	
	\begin{abstract}	
    Online social media platforms enable influencers to distribute content and quickly capture audience reactions, significantly shaping their promotional strategies and advertising agreements. Understanding how sentiment dynamics and emotional contagion unfold among followers is vital for influencers and marketers, as these processes shape engagement, brand perception, and purchasing behavior. While sentiment analysis tools effectively track sentiment fluctuations, dynamical models explaining their evolution remain limited, often neglecting network structures and interactions both among blogs and between their topic-focused follower groups. In this study, we tracked influential tech-focused Weibo bloggers over six months, quantifying follower sentiment from text-mined feedback. By treating each blogger’s audience as a single ``macro-agent'', we find that sentiment trajectories follow the principle of iterative averaging -- a foundational mechanism in many dynamical models of \emph{opinion formation}, a theoretical framework at the intersection of social network analysis and dynamical systems theory. The sentiment evolution aligns closely with opinion-dynamics models, particularly modified versions of the classical French–DeGroot model that incorporate delayed perception and distinguish between expressed and private opinions.  The inferred influence structures reveal interdependencies among blogs that may arise from homophily, whereby emotionally similar users subscribe to the same blogs and collectively shape the shared sentiment expressed within these communities.
    \end{abstract}

	\section{Introduction}

	Emotional contagion -- the transfer of emotional states between individuals -- has been extensively studied in social psychology and supported by numerous experiments~\cite{hatfield1993emotional, barsade2002ripple, Barsade2018, herrando2021emotional,kramer2014emotional}. Social media platforms, with their multimodal communication (text, images, video), enable rapid emotional diffusion, making them ideal for studying contagion. Emotions expressed in posts influence users, with homophily amplifying this effect by connecting emotionally similar individuals~\cite{rosenbusch2019multilevel} who often subscribe to the same blogs and channels, forming new communication pathways. Platforms like Facebook, YouTube, Weibo, and Twitter provide powerful tools for analyzing these dynamics~\cite{yin2021, yin2022, Meyer2023, Song2024}.
    
    As users gain popularity and become influencers, they become more sensitive to audience feedback, which shapes their self-promotion strategies and advertising agreements. Managing emotional contagion is crucial: understanding how emotions spread among followers and how sentiment evolves in response to content helps influencers sustain engagement, protect their public image, and optimize promotions. These dynamics are also central to online marketing and brand management, offering insights into how emotions drive consumer behavior and brand adoption~\cite{liu2022analysis}. Additionally, tracking emotional dynamics in public discourse -- especially during crises like the COVID-19 pandemic -- is vital for effective public health communication~\cite{crocamo2021surveilling,Ahmed2021}.

    \subsubsection*{Modeling Emotions via Sentiments and Opinions}

    In computational models of emotion, discrete states -- such as anger, fear, joy, or excitement -- are often represented as points in a two-dimensional~\cite{russell2003core} or three-dimensional~\cite{mehrabian1996pad} affective space, defined by valence (pleasure), arousal (activation), and optionally, dominance. This representation supports various dynamical models of emotional contagion, which describe how emotions evolve through interactions in social networks~\cite{vanHaeringen2023emotion}. As noted in~\cite{vanHaeringen2024}, applying and validating these models remains challenging even in small groups, due to technical and ethical constraints, especially given that emotions are inherently internal and short-lived~\cite{deonna2008,VendrellFerran2022}.
    
    In view of these challenges, emotional dynamics in large-scale social media are typically examined through observable fluctuations in \emph{sentiments}~\cite{Naskar2020,liu2022analysis} -- enduring attitudes or beliefs that accumulate and integrate transient emotional states. While there is no consensus on the exact definition of sentiment in philosophical and psychological literature~\cite{VendrellFerran2022}, in text processing the term generally refers to a scalar value representing the emotional tone of a text or the writer's attitude toward a specific issue, which can be positive, negative, or neutral.
    Specialized tools for sentiment extraction~\cite{nandwani2021review,Zhang2023survey,Xu2025_survey,Rimpy2024} make it possible to trace sentiment dynamics at scale in social media. From a modeling perspective, sentiment can be viewed as a special case of \emph{opinion}, as defined in sociology~\cite{Friedkin:2015}, sociophysics~\cite{Noorazar2020}, economics~\cite{Grabisch2023,mastroeni2019agent}, and systems theory~\cite{proskurnikov2018tutorial,anderson2020ifac}. However, research on sentiment dynamics and opinion formation has evolved in parallel, with limited integration between these fields.

    In the literature on sentiment dynamics modeling, the prevailing approach is macroscopic, emphasizing how sentiment distributions evolve across social groups over time. Representative examples include a model inspired by population games~\cite{franke2014}, a kinetics-based framework proposed by Rey~\cite{Rey2010}, hidden Markov models capturing the joint evolution of sentiment-topic distributions~\cite{he2014djst}, and compartmental models adapted from epidemiology~\cite{yin2021, yin2022, chen2025s3eir}.

    In a parallel line of research, opinion formation models have been actively developed over the past decades~\cite{Friedkin:2015,Noorazar2020,Grabisch2023,mastroeni2019agent,proskurnikov2018tutorial,Zhou2024fj,He2023,Bernardo2024survey,anderson2020ifac,ravazzi2021learning}. Beyond macroscopic approaches, this field includes a broad class of microscopic, or agent-based, models rooted in foundational works such as French’s study of social power~\cite{french1956formal} followed by DeGroot's model of rational consensus~\cite{degroot1974reaching} and Abelson's models of attitude dynamics~\cite{Abelson:1964}. Even relatively simple linear models -- such as the Friedkin–Johnsen model~\cite{Friedkin:2015,Zhou2024fj} -- can reproduce complex emergent behaviors observed in empirical settings and fit experimental data~\cite{FRIEDKINPROBULLO:2021,FriedProsk2019}. Tools from control theory and signal processing facilitate efficient parameter learning and prediction of long-term (asymptotic) behaviors~\cite{ravazzi2021learning,proskurnikov2018tutorial}.

    From a sociological perspective, the term ``opinion'' can be broadly understood as an individual's cognitive orientation toward an issue, person, or event~\cite{Friedkin:2015}. In contemporary opinion dynamics research, opinions are often modeled as numerical variables that represent social actors (individuals or groups) and evolve through social interactions. Sentiment scores, which reflect the emotional states of individuals and communities, can be interpreted as a continuous (real-valued) opinion,\footnote{Note that some works in sentiment analysis distinguish between sentiments and opinions, defining an opinion, for example, as a pair consisting of a sentiment (a real-valued score) and a discrete topic~\cite{Liu2015} or a discrete attribute determined by the sentiment scores~\cite{lian2022}.} situating them within the framework of opinion formation modeling in social groups.


	\subsubsection*{Merging Two Lines of Research: Objectives of This Work}

    Sentiments expressed by individual users or user groups (e.g., subscribers to different blogs) on social media can coevolve~\cite{zhang2021multilevel} due to social ties among users -- an effect that opinion dynamics models are well-equipped to capture. A natural question arises: Can agent-based opinion formation models -- traditionally focused on decision-making, and information spread rather than emotional contagion -- effectively capture sentiment trends, thereby offering a parsimonious framework for modeling emotional dynamics?
    
    This question motivates the present study. Modeling sentiment spread with parsimonious agent-based frameworks offers both practical and theoretical advantages. If simple models like the French–DeGroot or Friedkin–Johnsen models prove effective, they enable real-time, interpretable forecasting of collective mood shifts, valuable for marketers~\cite{liu2022analysis}, health authorities~\cite{Ahmed2021}, and crisis managers~\cite{Ilyas2025}. Learning influence structures also reveals key agents responsible for  sentiment shifts, supporting targeted interventions~\cite{He2023}.

    An early example of agent-based modeling is the seminal work by Macropol et al.~\cite{Macropol2013}, which applied an averaging-based model with time-varying activity-driven influence matrices to political sentiment on Twitter.\footnote{More recent work~\cite{kozitsin2022} further supports this view, demonstrating that sentiment updates on VKontakte tend to follow patterns consistent with iterative averaging.} Similarly, Das et al.~\cite{das2019} validated a probabilistic averaging-based model on Twitter data from vaccination debates, demonstrating its capacity to capture sentiment evolution.

	\subsubsection*{Contributions of This Work}	

	We analyze the sentiment dynamics of subscribers to several influential self-media blogs on Weibo, a leading Chinese microblogging platform. The platform offers convenient tools for tracking user sentiment on specific topics over extended periods; the typically short length of Weibo comments allows for relatively accurate sentiment analysis.

    The fan group of each influencer is modeled as a unified ``macro-agent'', since subscribers are exposed to the same content, topics, and interaction context. Aggregating their responses captures the group’s collective emotional orientation -- its ``opinion'' -- framing sentiment evolution within the opinion dynamics paradigm. This approach is supported by homophily~\cite{rosenbusch2019multilevel}, which promotes sentiment alignment among like-minded users, and by echo chamber effects~\cite{Song2024}, where repeated exposure to similar content reinforces emotional convergence. Both effects are further amplified by recommender algorithms~\cite{Bakshy2015} that prioritize content matching users’ prior preferences, increasing intra-group cohesion and reducing exposure to opposing views.

    To compute the collective opinion, we gather comment data from the fan groups of selected self-media bloggers, reflecting followers’ subjective sentiments toward the bloggers' posts on a specific topic of high visibility and longevity -- Huawei 
    products. Huawei is a prominent domestic tech brand that consistently generates high levels of user engagement and discussion on social media. We collect comments associated with these posts, extract primary comments (i.e., those responding directly to the blogger), and perform sentiment analysis as described in Section~\ref{sec.method}. The collective opinion is then computed by averaging the sentiment scores across followers. The time series describing sentiment evolution in each blog are then used to evaluate the performance and applicability of linear opinion formation models, as surveyed in Section~\ref{subsec:opinion}. The choice of these models is motivated by the observed averaging tendency in the data~-- specifically, the non-expanding range of opinions -- and by the availability of well-established methodologies for learning influence matrices~\cite{nematollahzadeh2020learning,ravazzi2021learning}. We show that while the basic DeGroot~\cite{degroot1974reaching} and Friedkin–Johnsen~\cite{Friedkin:2015} models fall short in capturing the dynamics of collective sentiment, more advanced formulations that incorporate delay effects and private (undisclosed) opinions~\cite{ye2019influence} can approximate these dynamics with sufficient accuracy.

	\subsubsection*{Organization of the Paper}

	Section~\ref{sec.method} presents the methodological and algorithmic aspects of our approach, including procedures for selecting the most popular bloggers on Weibo, quantifying the audience responses with sentiment analysis algorithms, and assessing sentiment dynamics using opinion dynamics models. Section~\ref{sec:results} 
    presents the main findings: a general tendency toward averaging in the observed data and a comparative evaluation of opinion formation models. Models that incorporate private opinions demonstrate superior performance in fitting the data compared to other classical opinion dynamics models and exhibit considerable predictive power.
     Section~\ref{sec:conclusions} concludes the paper. Technical details on sentiment analysis algorithms are provided in the Appendix.

    \section{Methods and Implementation Aspects} \label{sec.method}
	
	This section describes our methodology and its implementation, with a focus on data collection and the application of opinion formation models, including parameter estimation.
	
	\subsection{Influencers and the Followers' Reactions}	
	
	Weibo's user-centric design enables a wide range of individual viewpoints on diverse topics over various time periods. Users frequently share their sentiments and opinions in real time as events unfold, making it well-suited for observing sentiment dynamics over time. However, the platform’s broad and varied content poses challenges for isolating topic-specific posts and comments.
	
	To indicate a blogger’s popularity, Weibo assigns a \textbf{V} label to their profile and includes them in the V-influence list.\footnote{Ranking rules are available on\;
    \url{https://v6.bang.weibo.com/rule}} This curated list features 200 of the most influential verified bloggers in a given field. Rather than relying solely on follower count, Weibo ranks bloggers based on a proprietary ``popularity'' metric. The V-influence list (updated monthly) serves as a benchmark for identifying bloggers who lead trending discussions and engage millions of users daily.

    For our experiments, we selected the top 
    bloggers from the V-influence list, based on the following criteria: (i) each must be a media outlet or celebrity with a red \textbf{V} label; (ii) their content must focus on the digital domain (chosen for its topical neutrality); (iii) they must have at least 1,000,000 followers; and (iv) they must publish at least one Huawei-related post every 15 days, containing either the keyword \begin{CJK}{UTF8}{gbsn}华为\end{CJK} or Huawei. Our sentiment analysis is based on top-level comments rather than the bloggers' posts themselves. 
   
    Of the selected bloggers, only seven met all the specified criteria (Table~\ref{tab:bloggers}). We identified them from the V-influence list and crawled their blogs from May 16 to November 15, 2023,\footnote{This period spans twelve 15-day intervals, each covering either the 1st-15th or 16th-end of the month. Notably, the Weibo API provides access to historical data only for the past six months. Measuring sentiment too frequently (e.g., daily) risks capturing intervals with no Huawei-related posts, while overly infrequent sampling may not yield sufficient data for effective model training and validation.} using the pattern \texttt{https://weibo.com/u/\{blogger\_id\}}.
    
	\begin{table}[h!]
        \centering
		\caption{Selected Weibo bloggers, sorted by ID}
		\label{tab:bloggers}
		\begin{CJK}{UTF8}{gbsn}
			\tabcolsep=0.11cm
			\begin{tabular}{llll}
				\toprule
				\# & \bf Blogger & \bf Blogger ID & \bf Followers \\
				\midrule
				1 & 科技小辛 (Xiaoxin Technology) & 1595443924 & 1.611M \\ \rowcolor{LightGray}
				2 & 李杰灵 (Li Jieling) & 1783497251 & 2.982M \\
				3 & 小白测评 (Novice Evaluation) & 2022252207 & 3.470M \\  \rowcolor{LightGray}
				4 & 数码疯报 (Digital Crazy News) & 2561744167 & 2.656M \\
				5 & 定焦数码 (Fixed Focus Digital) & 5821279480 & 1.072M \\  \rowcolor{LightGray}
				6 & 勇气数码君 (Courage Digital Master) & 7109370363 & 1.708M \\
				7 & 白问视频 (Baiwen Video) & 7239083016 & 1.105M \\  
				\bottomrule
			\end{tabular}
		\end{CJK}
	\end{table}
		
	For each blog, Weibo provides access to a list of posts along with detailed metadata, including the number of likes,\footnote{The likes (thumbs-up emoji) is the only form of emotional feedback available on Weibo.} reposts, and comments (see Table~\ref{tab:posts}). We extract threads containing the previously specified Huawei-related keywords, focusing exclusively on \emph{top-level} comments. These comments capture users' direct emotional responses to the influencer's original posts, rather than interactions with other commenters.
	\begin{table*}[t]
        \centering
		\caption{An extract from a dataset 
			with posts by \begin{CJK}{UTF8}{gbsn}李杰灵\end{CJK} (Li Jieling)}
		\label{tab:posts}
		\begin{CJK}{UTF8}{gbsn}
			\begin{tabular}{llllll}
				\toprule
				\bf Post ID & \bf Post & \bf Date and time & \bf Likes & \bf Reposts & \bf Comments \\
				\hline
                \midrule
				\ldots & \ldots & \ldots & \ldots\\
				Msy8qEFGc & \#趣体\ldots一起探索华为\ldots & 2023-02-11 19:40&1227&3498&1081 \\  \rowcolor{LightGray}
				Mt0OgwLj3 & \#趣体\ldots我带着华为学\ldots& 2023-02-14 20:40&2647&6085&1423 \\ 
				MuEs3o7b2 & 兄弟们\ldots搞一个华为光\ldots & 2023-02-25 15:25&619&25&251 \\  \rowcolor{LightGray}
				Mv9TTmrT4 & \#趣看\ldots次还有华为的\ldots & 2023-02-28 23:29&195&197&72\\ 
				Mvx9b5qFR & \#谁是\ldots只蹭了华为的\ldots & 2023-03-03 10:40&3978&4865&1715 \\  \rowcolor{LightGray}
				MvH6a0RQz &\#趣体\ldots半的华为Ma\ldots & 2023-03-04 12:00&384&667&173 \\ 
				\ldots & \ldots & \ldots & \ldots \\
				\bottomrule
			\end{tabular}
		\end{CJK}
	\end{table*}
		
	\subsection{Sentiment Analysis}	
	
	Using a sentiment analysis algorithm, we compute a sentiment score for each of the selected comments, ranging from~$0$ (maximally negative) to~$1$ (maximally positive).
	
    Sentiment analysis algorithms, typically trained on datasets like movie and product reviews, often perform poorly on platforms like Weibo due to the abundance of slang and platform-specific expressions. To find the most suitable algorithm for this context, we evaluate several approaches,  including the Naive Bayes model for text classification~\cite{mosteller1964inference}, the BiLSTM model~\cite{schuster1997bidirectional,hochreiter1997long}, and a pre-trained BERT model~\cite{devlin2018bert}, specifically \emph{bert-base-chinese},\footnote{\url{https://huggingface.co/google-bert/bert-base-chinese}.} designed for processing Chinese texts. We fine-tune the algorithms on two publicly available sentiment annotation datasets.  Dataset~1 (\emph{ChnSentiCorp-Htl-ba-10000}\footnote{\url{https://ieee-dataport.org/open-access/chnsenticorp}.}) consists of hotel reviews scraped from \url{https://trip.com}, with labeled sentiments organized into subsets of varying sizes. Dataset~2 (\emph{weibo\_senti\_100k}~\cite{weibosenti}) comprises manually labeled positive and negative posts scraped from Weibo.

\begin{figure}[h]
		\centering
		\includegraphics[width=0.8\linewidth]{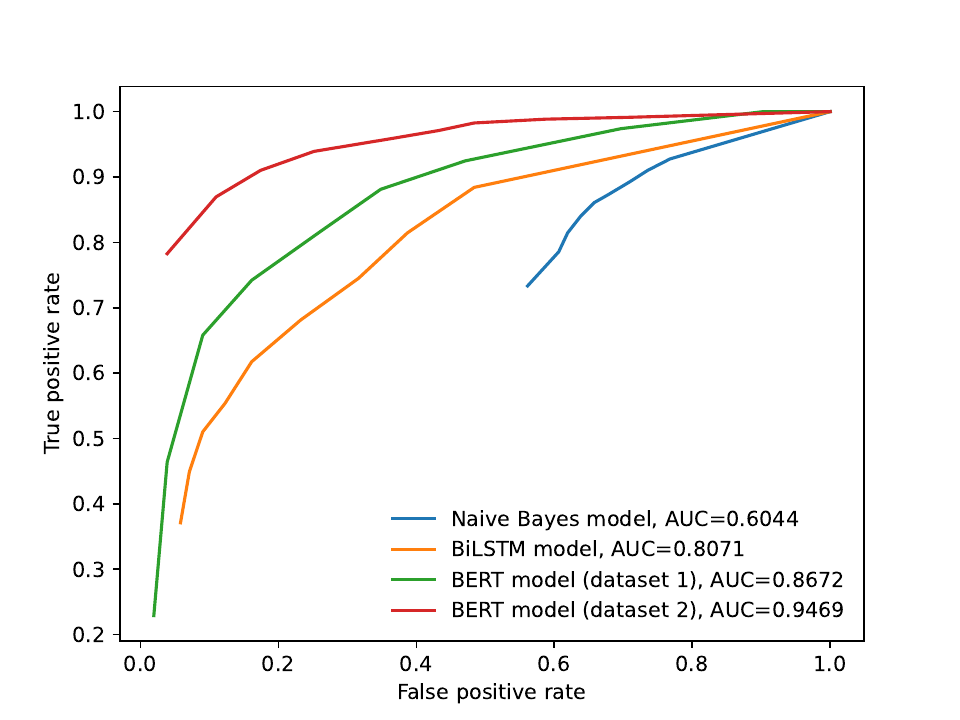}
		\caption{Sentiment analysis algorithms: ROC curves.}
		\label{fig:roc}
	\end{figure}

    To compare different models, we use a manually labeled 2018 Weibo test dataset (see Appendix for technical details). The BERT model fine-tuned on the \emph{weibo\_senti\_100k} consistently outperforms other approaches across standard metrics, making it the preferred choice for sentiment analysis on Weibo blogs (Table~\ref{tab:comparison}). Here, we report the Receiver Operating Characteristic (ROC) curves (Fig.~\ref{fig:roc}), which visualize the diagnostic ability of binary classifiers by plotting the true positive rate against the false positive rate at various classification thresholds. A higher area under the ROC curve (AUC), a standard metric for evaluating sentiment analysis algorithms, indicates superior overall performance across all classification thresholds.
    
	\subsection{Collective Follower Sentiment (Opinion)}\label{subsec:collective_sent}
	
    After computing sentiment scores for individual posts as described above, we define the collective \emph{opinion} of a blog's subscribers -- treated as a single social actor -- as the average sentiment across all posts. The procedure for computing collective opinion is formalized mathematically as follows.
	
	Let influencers be indexed by $\mathcal{B} = \{1, \ldots, B\}$, disseminating content on a specific topic over time periods\footnote{In our experiments, we set $B = 7$ and $T = 12$.} $\mathcal{T} = \{1, \ldots, T\}$.  Since the content analyzed is largely emotionally neutral (e.g., news about Huawei), the observed sentiment fluctuations are assumed to arise primarily from emotional contagion among followers -- either within a single blog or across multiple blogs. Followers' sentiment is evaluated by analyzing the comments left in response to the influencer's posts. As noted earlier, we focus exclusively on top-level comments, which respond directly to the influencer's original post rather than to other users.
	
    In the subsequent text, $\mathcal{P}_b(t)$ denotes the set of posts by blogger $b \in \mathcal{B}$ during period $t \in \mathcal{T}$. For each post $p \in \mathcal{P}_b(t)$, we identify the set of top-level comments $\mathcal{C}_{p,b}(t)$ associated with this post.  The Weibo platform supports only one type of emotional reaction: the thumbs-up. We denote by $\lambda_c(t)$ the number of thumbs-up votes on a top-level comment $c$ during period $t$, which serves as a measure of how much support the comment receives from other users.
    Similarly, $\lambda_p(t)$ denotes the number of thumbs-up reactions received by a post $p$ during period $t$, reflecting the perceived importance of the blogger's post within the Weibo audience.

    \textbf{Step 1.} Each top-level comment $c \in \mathcal{C}_{p,b}(t)$ is evaluated using a sentiment analysis algorithm (we use the BERT model due to its superior performance), which assigns a sentiment score $\sigma_c(t) \in [0,1]$. As discussed earlier, a sentiment score of 0 represents completely negative sentiment, while 1 corresponds to highly positive sentiment.

    \textbf{Step 2.} The average sentiment of followers, $\sigma_p(t) \in [0,1]$, toward a post $p \in \mathcal{P}_b(t)$ is defined as the weighted average of the sentiment scores expressed in top-level comments during the specified period. The weights are proportional to the number of thumbs-up reactions each comment receives: 
	\begin{equation*}\label{post-score}
		\sigma_p(t) = \frac{\sum_{c \in \mathcal{C}_{p,b}(t)} \lambda_c(t)\sigma_c(t)}{\sum_{c \in \mathcal{C}_{p,b}(t)} \lambda_c(t)}, \quad p \in \mathcal{P}_b(t),\; b \in \mathcal{B},\; t \in \mathcal{T}.
	\end{equation*}
    A similar approach is used by \cite{wong2016quantifying} to assess the political leanings of Twitter users during the 2012 U.S. election cycle, where a user's retweet approval score is calculated by aggregating the scores of their peers, weighted by the number of retweets they receive.
    
	\textbf{Step 3.} Next, we define the followers' sentiment regarding the entire topic-related content during period $t$ as the weighted average of post-level sentiments, where the weights are proportional to the number of thumbs-up reactions each post receives:
    \begin{equation*}\label{blogger-score}
		\sigma_b(t) = \frac{\sum_{p \in \mathcal{P}_b(t)} \lambda_p(t)\sigma_p(t)}{\sum_{p \in \mathcal{P}_b(t)} \lambda_p(t)}, \quad b \in \mathcal{B},\; t \in \mathcal{T}.
	\end{equation*}
	
	This score reflects followers' sentiment -- not the influencer's own views -- and is derived entirely from their feedback by aggregating top-level comments and then averaging across posts. By selecting prominent blogs with consistent topic coverage and high engagement,\footnote{Posts receive 60-70 top-level comments and 800+ likes on average.} and applying two-stage averaging, we reduce algorithm-induced noise in individual sentiment estimates.

\begin{figure}[htb]
        \small
		\centering
		\begin{tikzpicture}
			\begin{axis}
				[legend columns=4,
				legend style={at={(0.5,-0.25)},anchor=north},
				xlabel={Period, $t$},
				ylabel={Observed sentiments, $\sigma_b(t)$},
				xlabel near ticks,
				ylabel near ticks,
				xmin=1,
				xmax=12,
				ymin=0.38,
				ymax=0.85,
				xtick={1,2,3,4,5,6,7,8,9,10,11,12},
				ytick={0.4,0.5,0.6,0.7,0.8},
				width=8.5cm,
				height=6cm,
				legend cell align=left,
				legend style={draw=none, font=\scriptsize}
				]
				\pgfplotstableread[col sep=&, row sep=\\]{
					1 & 2 & 3 & 4 & 5 & 6 & 7 & 8 & 9 & 10 & 11 & 12 \\
                    0.542237 & 0.69269 & 0.719408 & 0.656396 & 0.662846 & 0.72706 & 0.582414 & 0.577169 & 0.658991 & 0.808328 & 0.775289 & 0.713721 \\
                    0.690464 & 0.685991 & 0.626356 & 0.593513 & 0.809609 & 0.665289 & 0.529654 & 0.678874 & 0.659458 & 0.682581 & 0.709356 & 0.636683 \\
                    0.640777 & 0.601049 & 0.617184 & 0.593332 & 0.643933 & 0.648383 & 0.601566 & 0.568535 & 0.610053 & 0.552651 & 0.600009 & 0.499918 \\
                    0.687817 & 0.623997 & 0.529495 & 0.621614 & 0.649233 & 0.708136 & 0.598221 & 0.604527 & 0.607142 & 0.600858 & 0.592896 & 0.627276 \\
                    0.473992 & 0.497273 & 0.513026 & 0.506612 & 0.546008 & 0.594167 & 0.523988 & 0.545223 & 0.571942 & 0.573838 & 0.484829 & 0.561077 \\
                    0.740787 & 0.669839 & 0.666616 & 0.742431 & 0.571651 & 0.73716 & 0.655859 & 0.636605 & 0.669664 & 0.710904 & 0.679262 & 0.784448 \\
                    0.551033 & 0.557094 & 0.537976 & 0.49017 & 0.632734 & 0.766128 & 0.574511 & 0.653399 & 0.694449 & 0.679584 & 0.680478 & 0.708864 \\
				}\datatable;
				\pgfplotstabletranspose\datatable{\datatable};
				\addplot[mark=o, blue!60!black, thick] table [x index=1, y index=2]{\datatable};\addlegendentry{$\sigma_1(t)\quad$}
				\addplot[mark=o, red!90!black, thick] table [x index=1, y index=3]{\datatable};\addlegendentry{$\sigma_2(t)\quad$}
				\addplot[mark=o, orange!80!yellow, thick] table [x index=1, y index=4]{\datatable};\addlegendentry{$\sigma_3(t)\quad$}
				\addplot[mark=o, green!60!black, thick] table [x index=1, y index=5]{\datatable};\addlegendentry{$\sigma_4(t)\quad$}
				\addplot[mark=o, magenta!90!orange, thick] table [x index=1, y index=6]{\datatable};\addlegendentry{$\sigma_5(t)\quad$}
				\addplot[mark=o, cyan!90!black, thick] table [x index=1, y index=7]{\datatable};\addlegendentry{$\sigma_6(t)\quad$}
				\addplot[mark=o, green!30!yellow, thick] table [x index=1, y index=8]{\datatable};\addlegendentry{$\sigma_7(t)\quad$}
			\end{axis}
		\end{tikzpicture}
		\caption{Observed collective sentiments in seven blogs.}
		\label{fig:observed}
	\end{figure}
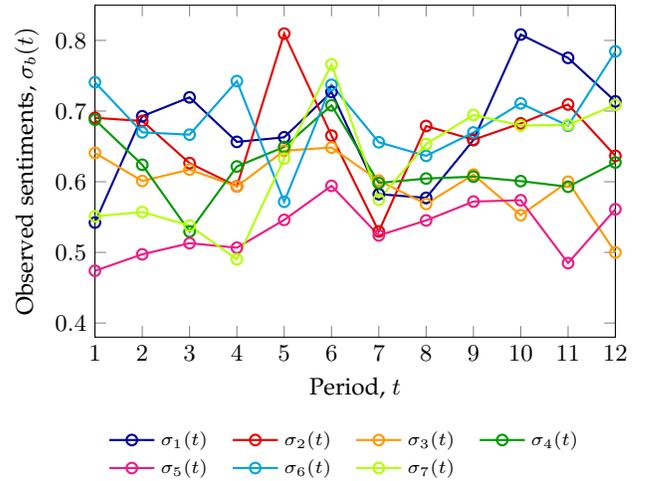

    Fig.~\ref{fig:observed} presents the evolution of collective sentiments, computed as described in Subsect.~\ref{subsec:collective_sent}, over twelve 15-day periods. The data show an overall positive\footnote{Recall that the scores of 0 and 1 correspond, respectively, to a maximally negative and a maximally positive audience attitudes.} attitude of the subscribers towards the bloggers' posts, with an average score of 0.6263 across all bloggers and time intervals.

\subsection{Opinion Dynamics Models}\label{subsec:opinion}
	
Assigning the sentiment score $\sigma_b(t)$ to blog $b$ links sentiment analysis to opinion dynamics by treating the blog -- comprising an influencer and their audience -- as a collective social actor. We interpret this sentiment 
$\sigma_b(t)$ as the actor's collective opinion  at period $t$. Below, we provide a brief description of the models under consideration that are summarized (in a compact matrix form) in Table~\ref{tab:allmodelsNew}.
   	
The models operate on a discrete time scale, aligning naturally with observations spaced at fixed time intervals.

\subsubsection*{Iterative Averaging of the Sentiments}

While there is no widely accepted agent-based model for sentiment dynamics in the literature, we begin by testing the most basic class of opinion formation models, rooted in the well-known French--DeGroot (FDG) and Friedkin--Johnsen (FJ) frameworks~\cite{french1956formal,degroot1974reaching,Friedkin:2015,ravazzi2021learning}. 

\textbf{The FDG model:} At each step, an agent's updated opinion is a (weighted) average of the current opinions: 
\begin{equation}\label{eq:fdg}
x_b(t+1) = \sum\nolimits_{k \in \mathcal{B}} w_{bk}x_k(t),\;\; b\in\mathcal{B},
\end{equation}
starting from some initial values $x_b(1)$. Here the weights $w_{bk}$ form a row-stochastic matrix\footnote{That is, all entries of the matrix are nonnegative $w_{bk}\geq 0$, and each of its rows sums to 1.} $W$, reflecting the relative influence of agent $k$ on agent $b$. The FDG model serves as a basic framework for multi-agent consensus, as opinions ultimately converge in generic scenarios, e.g., when $W$ is irreducible and aperiodic~\cite{degroot1974reaching} (more generally, when its graph has a directed spanning tree~\cite{proskurnikov2017tutorial}). Hence, the FDG model is insufficient to capture social dynamics, exhibiting persistent disagreement is common over long time intervals.

\textbf{The FJ model:} The FJ model extends the FDG framework by introducing agents who are ``anchored'' to \emph{innate} opinions\footnote{In the original FJ model~\cite{FJ99}, the innate opinion $z_b$ of agent $b$ is typically taken as their initial opinion, which preserves some memory of past influences on the group. In contrast, unlike in experiments validating the FJ model~\cite{FJ99,FriedProsk2019,FRIEDKINPROBULLO:2021}, we do not observe sentiment dynamics from the inception of the blog; consequently, the true initial opinions cannot be directly measured.} $z_b$, incorporating these fixed predispositions into every step of the opinion update process:
\begin{equation}\label{eq:fj}
x_b(t+1) = s_b\sum\nolimits_{k \in \mathcal{B}} w_{bk}x_k(t) + (1-s_b)z_b,\;\;b\in\mathcal{B}.
\end{equation}

The coefficient $s_b \in [0,1]$ represents an agent's susceptibility to social influence, while the complementary quantity $1 - s_b$ captures their anchorage to innate opinion, keeping their individual memory. In generic scenarios, e.g., when the influence matrix $W$ is irreducible and at least one agent retains memory ($s_b < 1$) -- the opinion vector converges to a unique equilibrium, which generally exhibits disagreement or polarization rather than consensus ~\cite{Friedkin:2015}.

Comparing the FDG and FJ models, one observes that the former serves as a deterministic analogue of a Markov chain: each agent's next opinion depends solely on previous opinions. The FJ model, by contrast, allows agents to retain memory of their innate opinions $z_b$ indefinitely, introducing persistent influence absent in the FDG framework. The dynamics of sentiments plausibly exhibit memory, as further supported by the analysis of opinion range dynamics in the next section. First, some users' reactions may be delayed, as comments are often formed in response to posts or comments from earlier periods. Consequently, the computed collective sentiment reflects users' attitudes toward the blogger's past activity rather than recent content alone. Second, Weibo's recommendation algorithm analyzes users' past behavior and promotes content accordingly, reinforcing sentiment persistence over time. However, in rapidly evolving social media -- where blogs can gain and lose millions of subscribers -- it is not obvious whether collective sentiments can remain indefinitely anchored to fixed memories.
    
    \textbf{The FDGM model:} For this reason, we explore an extension of the FDG model~\eqref{eq:fdg} that allows agents to retain memory, but only over a finite number of steps. This model, termed FDGM (the French--DeGroot model with memory), resembles the structure of the FJ model; however, the static innate opinion is replaced by the agent's previous opinion:
    \begin{equation}\label{eq:fj}
    x_b(t+1) = s_b\sum\limits_{k \in \mathcal{B}} w_{bk}x_k(t) + (1-s_b)x_b(t-\tau),\;\; b\in\mathcal{B}.
    \end{equation}

    
	
	\subsubsection*{Two-Layer Averaging Models}

    Models of opinion dynamics derived from the FDG framework typically assume that agents receive the exact opinions of their peers. Recently, models have been proposed that relax this assumption by incorporating a discrepancy between private and expressed opinions. In these models, individuals maintain private beliefs that remain hidden, while their publicly expressed opinions reflect both personal convictions and social pressures, such as conformity to perceived public opinion or a desire for social approval, as demonstrated in the Asch conformity experiments~\cite{ye2019influence,zhao2024opinion}.

    Although the opinion in our experiment represents not an individual view but the collective sentiment of a blog's audience, one may argue that publicly expressed sentiments still do not fully capture users' true opinions, being shaped instead by public influence -- for example, liking or reposting a post simply because many others have already done so (``bandwagon behavior''). In view of this, we consider the EPO (expressed-private opinion) model from~\cite{ye2019influence}, which naturally extends the FDG framework by assigning each ``macro-agent'' two opinions: a private opinion $x_b(t)$ representing the unobserved genuine sentiment of the blog's audience, and an expressed opinion $x_b^e(t) \equiv \sigma_b(t)$ identified with the observed collective sentiment.

    \textbf{The EPO model:} 
    The EPO model from~\cite{ye2019influence} follows the structure of the FJ model, allowing the agents' anchorage to static innate opinions $z_b$. The private and expressed opinions of each agent $b \in \mathcal{B}$ evolve as follows
	\begin{equation}\label{eq:epo}
    \begin{aligned}
		x_b(t+1) &= s_bw_{bb}x_b(t) + s_b\sum_{k\neq b} w_{bk} x^e_k(t)+(1-s_b)z_b,\\
		x^e_b(t) &= \varphi_b x_b(t) + (1-\varphi_b)\sum_{k\neq b} a_{bk}x^e_k(t-1).
	\end{aligned}
    \end{equation}
    Compared to the FJ model, the EPO model involves two additional parameters: the coefficients $\varphi_b\in [0,1]$ and the matrix $A$. When $\varphi_b=1$ for all $b$, the second equation in (\ref{eq:epo}) becomes redundant, and the EPO model reduces to the standard FJ model. In general, the second equation describes the ``visible'' (expressed) layer of opinion formation: the expressed opinion of each agent arises as a weighted average of their genuine (private) opinion and the expressed opinions of their peers. In our context, where peers represent other blogs in the network, the coefficient $\varphi_b$ regulates the extent to which the blog's audience expresses their authentic sentiment, while $1-\varphi_b$ measures the extent to which they conform to the opinions of the others. The expressed opinion of agent $k$ influences peer $b$ (where $b \ne k$) in two ways: through weight $w_{bk}$ in the private opinion dynamics (the ``hidden'' layer of opinion formation) and through weight $a_{bk}$ in the expressed opinion dynamics.   

    As discussed in~\cite{ye2019influence}, the matrices $A$ and $W$ are likely to encode the same influence graph and should therefore be compatible, in the sense that $a_{bk} > 0$ (for $b \ne k$) if and only if $w_{bk} > 0$. Rather than enforcing this equivalence directly -- which would introduce numerous nonlinear constraints -- we impose the structural restriction $W = D + (I-D)A$, where $D = \text{diag}(d_{11}, \ldots, d_{BB})$ with entries $d_{bb} \in [0,1]$. This parameterization has a natural interpretation: the influence weights $w_{bk}$ and $a_{bk}$ are proportional, with $w_{bk} = (1-d_b)a_{bk}$ for $b \ne k$, where $1-d_b$ serves as the proportionality coefficient. When $d_{bb} < 1$, the compatibility condition $w_{bk} > 0 \Longleftrightarrow a_{bk} > 0$ holds automatically for all pairs $b\ne k$. The case $d_{bb} = 1$ corresponds to an ``internally stubborn'' agent whose private opinion is decoupled from peer influence: 
    $x_b(t+1)=s_bx_b(t)+(1-s_b)z_b$.

    \textbf{The Reduced EPO Model:} 
    As discussed above, in the rapidly changing Weibo environment where influencers acquire and lose millions of subscribers, the relevance of long-term memory -- represented by innate opinions $z_b$ -- becomes uncertain.
    In view of this, we consider a reduced form of the EPO model, corresponding to $s_b = 1$. While the full EPO model extends the classical FJ framework (and reduces to it when $\varphi_b = 1$), the reduced EPO generalizes the FDG model.
    
    \textbf{The EPO model with time lag:} In a similar spirit to the FDGM models, we consider modifications of both the full and reduced EPO models that incorporate delayed expressed opinions (time lag), replacing the vector $x^e(t)$ with $x^e(t - \tau)$, where $\tau \geq 1$.
    \begin{table*}[t]
        \centering
		\caption{The opinion dynamics models under consideration (in experiments, we test lags $\tau=1$ and $\tau=2$).}
		\label{tab:allmodelsNew}
		\small
		\begin{tabular}{>{\raggedright\arraybackslash}m{2.2cm}|l|l}
			\toprule
			\bf Model Name & \bf Dynamics and Variables & \bf Parameters and Constraints \\
			\midrule\midrule
			FDG~model &
			$\begin{aligned}
				x(t+1) &= W x(t),\\
                x_b&=\sigma_b: \text{sentiment in blog $b\in\mathcal{B}$}
			\end{aligned}$ &
			$\begin{aligned}
				W: \text{row-stochastic matrix}
			\end{aligned}$\\ 
			\midrule
			FJ~model &
			$\begin{aligned}
				x(t+1) &= SWx(t) + (I-S)z,	\\
                x_b&=\sigma_b: \text{sentiment in blog $b\in\mathcal{B}$}
			\end{aligned}$ &
			$\begin{aligned}
				W &: \text{row-stochastic matrix},\\				
				S  &: \text{diagonal matrix with values in}\,[0,1],\\			
				z  &: \text{constant vector with values in}\,[0,1]\\			
			\end{aligned}$\\
			\midrule
			EPO~model &
			$\begin{aligned}
				x(t+1) &= S\left(\textrm{diag}(W) x(t) + (W-\textrm{diag}(W)) x^e(t)\right) \\
                    &\qquad + (I-S)z,\\
				x^e(t) &= \Phi x(t) + (I-\Phi)A x^e(t-1), \\
                x_b^e&=\sigma_b: \text{sentiment in blog $b\in\mathcal{B}$}\\
                x_b&: \text{private opinion in the blog (hidden)}
			\end{aligned}$
			&
			$\begin{aligned}
				W, A &: \text{nonnegative row-stochastic matrices},\\
				S, \Phi, D &: \text{diagonal matrices with values in } [0,1]\\
				&\quad \text{with } W = D + (I-D)A,\\
				z &: \text{constant vector with values in } [0,1]
			\end{aligned}$\\ 
            \midrule
			Reduced EPO~model ($S=I$) &
			$\begin{aligned}
				x(t+1) &= \textrm{diag}(W) x(t) + (W-\textrm{diag}(W)) x^e(t), \\
				x^e(t) &= \Phi x(t) + (I-\Phi)A x^e(t-1), \\
                x_b^e, x_b&: \text{same as in the EPO model}
			\end{aligned}$
			&
			$\begin{aligned}
				W, A&: \text{nonnegative row-stochastic matrices},\\
				\Phi, D &: \text{diagonal matrices with values in } [0,1]\\
				&\quad \text{with } W = D + (I-D)A
			\end{aligned}$\\
			\midrule
			FDGM~model with lag~$\tau$ &
			$\begin{aligned}
				x(t+1) &= SWx(t) + (I-S)x(t-\tau),\\
				x_b&=\sigma_b: \text{sentiment in blog $b\in\mathcal{B}$}
			\end{aligned}$ &
			$\begin{aligned}
				W &: \text{nonnegative row-stochastic matrix},\\
				S &: \text{diagonal matrix with values in } [0,1]
			\end{aligned}$\\ 
			\midrule
			EPO~model with lag~$\tau$ &
			$\begin{aligned}
				x(t+1) &= S\left(\textrm{diag}(W) x(t) + (W-\textrm{diag}(W)) x^e(t)\right) \\
                    &\qquad + (I-S)z,\\
				x^e(t) &= \Phi x(t) + (I-\Phi)A x^e(t-\tau-1), \\
                x_b^e, x_b &: \text{same as in the EPO model}
			\end{aligned}$ &
			same as in the EPO model\\
            \midrule
			Reduced EPO~model with lag~$\tau$
            &
			$\begin{aligned}
				x(t+1) &= \textrm{diag}(W) x(t) + (W-\textrm{diag}(W)) x^e(t), \\
				x^e(t) &= \Phi x(t) + (I-\Phi)A x^e(t-\tau-1),\\
                x_b^e, x_b &: \text{same as in the EPO model}
			\end{aligned}$ &
			same as in the reduced EPO model\\
			\bottomrule
		\end{tabular}
	\end{table*}

\subsection{Parameter Estimation}\label{subsec:ls}
    
	As is common in the identification of continuous-opinion models~\cite{ravazzi2021learning,Wai2016}, we employ standard least-squares regression to estimate the unknown parameters. The procedure is straightforward for the FDG and FJ models, where all opinions are observed. For instance, for the FDG model, we we minimize over the set of stochastic matrices $W$ the sum\footnote{The dataset covers $T=12$ periods, of which the first $T_{\mathrm{est}} = 10$ are used for estimation and the final two for performance assessment.} 
    \[
    \sum\nolimits_{t=1}^{T_{\mathrm{est}}-1} \|x(t+1) - W x(t)\|^2.
    \]

    Parameter estimation in the EPO models is less straightforward, as it also involves the unobserved vectors of private opinions $x(t)$. We therefore treat $x(t)$ as additional unknowns to be determined via least squares, along with the parameter matrices. For instance, in the reduced ($S = I$) EPO model, these unknowns are determined from the following optimization problem:
	\begin{align*}
		&\min_{W,A,\Phi,x(\cdot)}  \sum\nolimits_{t =1}^{T_{\mathrm{est}}-1} J_t,\\ 
        &\quad J_t=\|x(t+1) - \textrm{diag}(W) x(t) \nonumber- (W-\textrm{diag}(W)) x^e(t)\|^2 \nonumber \\
		&\quad\qquad + \|x^e(t+1) - \Phi x(t+1) - (I-\Phi)A x^e(t) \|^2. 
	\end{align*}
    Here $A$ is a row-stochastic matrix with zero diagonal ($a_{bb}=0$ for all $b$), and $W = D + (I-D)A$, where $D$ and $\Phi$ are diagonal matrices with entries in $[0,1]$. Since the opinions represent normalized sentiments, we constrain the private opinions component-wise to $[0,1]$, i.e., $x_b(t) \in [0,1]$ for all $b$ and $t$. In the case of the full EPO model, the matrix $S$ and the innate opinion vector $z$ are also included among the parameters to be estimated (see Table~\ref{tab:allmodelsNew}).

    Finally, the time lag $\tau$ in the FDGM and EPO models further reduces the number of available observations for estimation, with the summation starting from $t=1+\tau$.
    	
	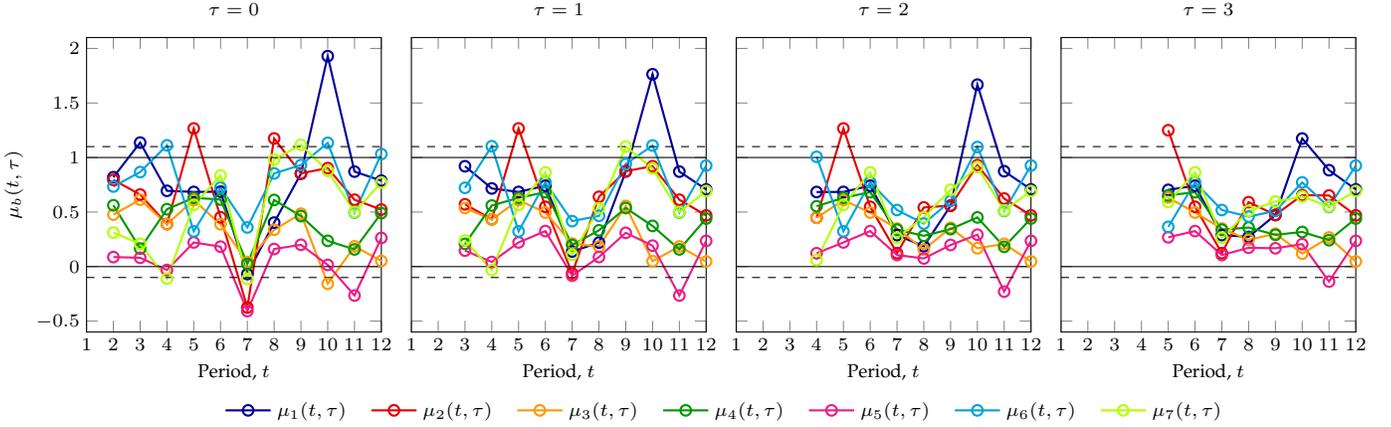
\begin{figure*}[htb]
		\scriptsize
		\centering
		\begin{tikzpicture}
                \begin{groupplot}[
                        group style={
                        group size=4 by 1,
                        horizontal sep=0.4cm,
                        y descriptions at=edge left,
                        group name=my plots
                    },
                    xlabel={Period, $t$},
                    ylabel={$\mu_b(t,\tau)$},
                    xmin=1,
                    xmax=12,
                    ymin=-0.6,
                    ymax=2.1,
                    xtick={1,2,3,4,5,6,7,8,9,10,11,12},
                    ytick={-0.5,0,0.5,1,1.5,2},
                    width=5.5cm,
                    height=5.5cm,
                    legend columns=7,
                    legend cell align=center,
                    legend style={
                        draw=none,
                        font=\scriptsize,
                        cells={anchor=center},
                        at={(2.2,-0.2)},
                        anchor=north
                    },
                    xlabel near ticks,
                    ylabel near ticks
                ]
                \nextgroupplot[title={$\tau=0$}]
						\pgfplotstableread[col sep=&, row sep=\\]{
						1 & 2 & 3 & 4 & 5 & 6 & 7 & 8 & 9 & 10 & 11 & 12 \\
						& 0.819721 & 1.13673 & 0.694686 & 0.684512 & 0.686841 & -0.0683437 & 0.403284 & 0.85123 & 1.92956 & 0.870779 & 0.788034 \\
						& 0.794612 & 0.660552 & 0.389993 & 1.2663 & 0.452507 & -0.375161 & 1.17453 & 0.854728 & 0.90312 & 0.612904 & 0.522804 \\
						& 0.476234 & 0.613618 & 0.389117 & 0.609538 & 0.388371 & 0.043027 & 0.337806 & 0.485069 & -0.157465 & 0.185227 & 0.0519489 \\
						& 0.562248 & 0.164886 & 0.526151 & 0.630548 & 0.615053 & 0.0235739 & 0.610741 & 0.46329 & 0.236038 & 0.157404 & 0.49042 \\
						& 0.0872619 & 0.0806087 & -0.0310778 & 0.22135 & 0.182695 & -0.408109 & 0.161029 & 0.199914 & 0.0154763 & -0.265266 & 0.262507 \\
						& 0.734072 & 0.866573 & 1.11156 & 0.323003 & 0.725158 & 0.358757 & 0.853995 & 0.931087 & 1.13432 & 0.495199 & 1.03153 \\
						& 0.311483 & 0.20829 & -0.110742 & 0.565144 & 0.835051 & -0.114306 & 0.981344 & 1.11654 & 0.878658 & 0.499954 & 0.771312 \\
					}\datatable;
					\pgfplotstabletranspose\datatable{\datatable};
					\addplot[mark=o, blue!60!black, thick] table [x index=1, y index=2]{\datatable};\addlegendentry{$\mu_1(t,\tau)\quad$}
					\addplot[mark=o, red!90!black, thick] table [x index=1, y index=3]{\datatable};\addlegendentry{$\mu_2(t,\tau)\quad$}
					\addplot[mark=o, orange!80!yellow, thick] table [x index=1, y index=4]{\datatable};\addlegendentry{$\mu_3(t,\tau)\quad$}
					\addplot[mark=o, green!60!black, thick] table [x index=1, y index=5]{\datatable};\addlegendentry{$\mu_4(t,\tau)\quad$}
					\addplot[mark=o, magenta!90!orange, thick] table [x index=1, y index=6]{\datatable};\addlegendentry{$\mu_5(t,\tau)\quad$}
					\addplot[mark=o, cyan!90!black, thick] table [x index=1, y index=7]{\datatable};\addlegendentry{$\mu_6(t,\tau)\quad$}
					\addplot[mark=o, green!30!yellow, thick] table [x index=1, y index=8]{\datatable};\addlegendentry{$\mu_7(t,\tau)\quad$}
					\addplot[mark=none, black, domain = 1:12]{0};
					\addplot[mark=none, black, domain = 1:12]{1};
					\addplot[mark=none, black, domain = 1:12, dashed]{-0.1};
					\addplot[mark=none, black, domain = 1:12, dashed]{1.1};
                
                \nextgroupplot[title={$\tau=1$}]
						\pgfplotstableread[col sep=&, row sep=\\]{
						1 & 2 & 3 & 4 & 5 & 6 & 7 & 8 & 9 & 10 & 11 & 12 \\
						 & & 0.919866 & 0.716337 & 0.684512 & 0.741582 & 0.138111 & 0.219631 & 0.871626 & 1.76313 & 0.870779 & 0.707552 \\
						 & & 0.571089 & 0.433251 & 1.2663 & 0.548209 & -0.0620432 & 0.639655 & 0.874645 & 0.920467 & 0.612904 & 0.469411 \\
						 & & 0.536712 & 0.432437 & 0.609538 & 0.495284 & 0.210764 & 0.183972 & 0.555667 & 0.0497773 & 0.185227 & 0.0466434 \\
						 & & 0.208034 & 0.559753 & 0.630548 & 0.682341 & 0.198074 & 0.332614 & 0.536874 & 0.372824 & 0.157404 & 0.440334 \\
						 & & 0.146304 & 0.0420392 & 0.22135 & 0.32556 & -0.083537 & 0.0876977 & 0.309607 & 0.191753 & -0.265266 & 0.235697 \\
						 & & 0.721991 & 1.10365 & 0.323003 & 0.7732 & 0.416732 & 0.465091 & 0.940535 & 1.11027 & 0.495199 & 0.926183 \\
						 & & 0.239825 & -0.0319756 & 0.565144 & 0.863884 & 0.108127 & 0.534447 & 1.10056 & 0.900384 & 0.499954 & 0.692539 \\
					}\datatable;
					\pgfplotstabletranspose\datatable{\datatable};
					\addplot[mark=o, blue!60!black, thick] table [x index=1, y index=2]{\datatable};
					\addplot[mark=o, red!90!black, thick] table [x index=1, y index=3]{\datatable};
					\addplot[mark=o, orange!80!yellow, thick] table [x index=1, y index=4]{\datatable};
					\addplot[mark=o, green!60!black, thick] table [x index=1, y index=5]{\datatable};
					\addplot[mark=o, magenta!90!orange, thick] table [x index=1, y index=6]{\datatable};
					\addplot[mark=o, cyan!90!black, thick] table [x index=1, y index=7]{\datatable};
					\addplot[mark=o, green!30!yellow, thick] table [x index=1, y index=8]{\datatable};
					\addplot[mark=none, black, domain = 1:12]{0};
					\addplot[mark=none, black, domain = 1:12]{1};
					\addplot[mark=none, black, domain = 1:12, dashed]{-0.1};
					\addplot[mark=none, black, domain = 1:12, dashed]{1.1};
							
                \nextgroupplot[title={$\tau=2$}]
						\pgfplotstableread[col sep=&, row sep=\\]{
						1 & 2 & 3 & 4 & 5 & 6 & 7 & 8 & 9 & 10 & 11 & 12 \\
						 & & & 0.683687 & 0.684512 & 0.741582 & 0.288769 & 0.186196 & 0.55754 & 1.66806 & 0.874428 & 0.707552 \\
						 & & & 0.447988 & 1.2663 & 0.548209 & 0.123602 & 0.542279 & 0.559471 & 0.930374 & 0.623833 & 0.469411 \\
						 & & & 0.447311 & 0.609538 & 0.495284 & 0.348723 & 0.155965 & 0.355435 & 0.168151 & 0.20823 & 0.0466434 \\
						 & & & 0.553315 & 0.630548 & 0.682341 & 0.338251 & 0.281979 & 0.343414 & 0.450954 & 0.181193 & 0.440334 \\
						 & & & 0.122264 & 0.22135 & 0.32556 & 0.105865 & 0.0743473 & 0.198042 & 0.29244 & -0.229544 & 0.235697 \\
						 & & & 1.00616 & 0.323003 & 0.7732 & 0.518687 & 0.394289 & 0.601618 & 1.09653 & 0.509451 & 0.926183 \\
						 & & & 0.0606389 & 0.565144 & 0.863884 & 0.264027 & 0.453087 & 0.703978 & 0.912793 & 0.514071 & 0.692539 \\
					}\datatable;
					\pgfplotstabletranspose\datatable{\datatable};
					\addplot[mark=o, blue!60!black, thick] table [x index=1, y index=2]{\datatable};
					\addplot[mark=o, red!90!black, thick] table [x index=1, y index=3]{\datatable};
					\addplot[mark=o, orange!80!yellow, thick] table [x index=1, y index=4]{\datatable};
					\addplot[mark=o, green!60!black, thick] table [x index=1, y index=5]{\datatable};
					\addplot[mark=o, magenta!90!orange, thick] table [x index=1, y index=6]{\datatable};
					\addplot[mark=o, cyan!90!black, thick] table [x index=1, y index=7]{\datatable};
					\addplot[mark=o, green!30!yellow, thick] table [x index=1, y index=8]{\datatable};
					\addplot[mark=none, black, domain = 1:12]{0};
					\addplot[mark=none, black, domain = 1:12]{1};
					\addplot[mark=none, black, domain = 1:12, dashed]{-0.1};
					\addplot[mark=none, black, domain = 1:12, dashed]{1.1};
			
                \nextgroupplot[title={$\tau=3$}]
				    \pgfplotstableread[col sep=&, row sep=\\]{
						1 & 2 & 3 & 4 & 5 & 6 & 7 & 8 & 9 & 10 & 11 & 12 \\
						 & & & & 0.703525 & 0.741582 & 0.288769 & 0.27235 & 0.472664 & 1.17428 & 0.883806 & 0.707552 \\
						 & & & & 1.25025 & 0.548209 & 0.123602 & 0.590736 & 0.474301 & 0.654963 & 0.651926 & 0.469411 \\
						 & & & & 0.63307 & 0.495284 & 0.348723 & 0.245319 & 0.301326 & 0.118374 & 0.267361 & 0.0466434 \\
						 & & & & 0.652814 & 0.682341 & 0.338251 & 0.357993 & 0.291135 & 0.317462 & 0.242343 & 0.440334 \\
						 & & & & 0.268277 & 0.32556 & 0.105865 & 0.172342 & 0.167893 & 0.205872 & -0.137719 & 0.235697 \\
						 & & & & 0.363804 & 0.7732 & 0.518687 & 0.458413 & 0.510032 & 0.771934 & 0.546086 & 0.926183 \\
						 & & & & 0.591352 & 0.863884 & 0.264027 & 0.510986 & 0.59681 & 0.642587 & 0.550361 & 0.692539 \\
					}\datatable;
					\pgfplotstabletranspose\datatable{\datatable};
					\addplot[mark=o, blue!60!black, thick] table [x index=1, y index=2]{\datatable};
					\addplot[mark=o, red!90!black, thick] table [x index=1, y index=3]{\datatable};
					\addplot[mark=o, orange!80!yellow, thick] table [x index=1, y index=4]{\datatable};
					\addplot[mark=o, green!60!black, thick] table [x index=1, y index=5]{\datatable};
					\addplot[mark=o, magenta!90!orange, thick] table [x index=1, y index=6]{\datatable};
					\addplot[mark=o, cyan!90!black, thick] table [x index=1, y index=7]{\datatable};
					\addplot[mark=o, green!30!yellow, thick] table [x index=1, y index=8]{\datatable};
					\addplot[mark=none, black, domain = 1:12]{0};
					\addplot[mark=none, black, domain = 1:12]{1};
					\addplot[mark=none, black, domain = 1:12, dashed]{-0.1};
					\addplot[mark=none, black, domain = 1:12, dashed]{1.1};
		      \end{groupplot}
        \end{tikzpicture}
		\caption{Functions $\mu_b(t,\tau)$ under two scenarios: no delay in averaging ($\tau = 0$) and with a delay in averaging for the first three lags ($\tau = 1,2,3$). The dashed lines indicate a 10\% exceedance or shortfall of the admissible values.}
		\label{fig:mu}
	\end{figure*}
    
\section{The Results}\label{sec:results}

Before fitting the opinion dynamics models to data, we notice one principal obstacle to applying the FDG model~\eqref{eq:fdg}, or any similar model that replaces $W$ by a time-varying matrix (e.g., the Hegselmann--Krause model or gossip-based models~\cite{proskurnikov2018tutorial}). Namely, the next-step opinions are weighted averages of current ones. Consequently, the opinion range (minimum-maximum) cannot expand, e.g., all opinions for $t > 1$ must remain within the range observed at $t = 1$. A glance at Fig.~\ref{fig:observed} reveals that even this latter condition is violated: the collective sentiments of the audiences of blogs 1, 2, and 7 at some periods fall outside the range of initially observed sentiments. We begin by computing the \emph{range violation indices} that quantify deviations of the opinions from the previous-step range.

\subsection{Range Violation Indices}

To quantify deviations from the averaging rule, we compare $\sigma_b(t)$ with the previous period's range $[m(t-1),M(t-1)]$, where
$m(t-1)=\min_{k\in\mathcal{B}} \sigma_k(t-1)$ and $M(t-1)=\max_{k\in\mathcal{B}} \sigma_k(t-1)$ are the minimum and maximum opinions at the previous period. We define the \emph{range violation index} as
\[
\mu_b(t,0):=\frac{\sigma_b(t)-m(t-1)}{M(t-1)-m(t-1)},\;\;t\geq 1.
\]

The dynamics of the variables $\mu_b(t,0)$ are shown in Fig.~\ref{fig:mu} (leftmost plot). The solid lines indicate the expected range $[0,1]$ (i.e., $m(t-1) \leq \sigma_b(t) \leq M(t-1)$). As can be seen, the sentiments of all blogs except blog 4 exceed this range by 10\% or more at some steps (the violation is marked by the dashed lines, corresponding to values of $-0.1$ and $1.1$).

One explanation for this behavior, suggested by the observed data, is that the opinion formation process involves a \emph{time lag}. Recall that an opinion here represents the collective sentiment of a blog's audience, reflecting the reactions of all its subscribers. Although a 15-day sampling period may seem sufficient, it is plausible that a significant fraction of subscribers respond with delay. Moreover, since posts and comments within each 15-day period are temporally dispersed, some content appearing near the end of one period may exert its full influence not in the current, but only in subsequent periods, leading to opinions that appear to violate the instantaneous averaging rule.

Extending the interval $[m(t),M(t)]$ to an opinion range that includes the previous $\tau\geq 1$ periods, we define
\begin{align*}
 m(t,\tau) &= \min_{0 \leq h \leq \tau}\min_{b \in \mathcal{B}} \sigma_b(t-h),\\ 
 M(t,\tau) &= \max_{0 \leq h \leq \tau}\max_{b \in \mathcal{B}} \sigma_b(t-h), \;\;t>\tau, 
 \end{align*}
 and computing the corresponding range violation indices,
 \begin{align*}		
		\mu_b(t,\tau) = \frac{\sigma_b(t) - m(t-1,\tau)}{M(t-1,\tau) - m(t-1,\tau)}, \quad t > \tau, \quad b \in \mathcal{B},
	\end{align*} 
one may observe (Fig.~\ref{fig:mu}) that most opinions satisfy the condition $m(t,\tau) \leq \sigma_b(t) \leq M(t,\tau)$ (equivalently, $\mu_b(t,\tau)\in[0,1]$) for $\tau=2$ (two-step delay), and deviate by no more than 10\% from this range when $\tau=1$ (one-step delay).

On the other hand, time lag is not the only possible explanation. Some sentiments leave the extended range at certain steps, and this effect persists even as the delay increases. This phenomenon may be explained by hidden (unobservable) variables, such as the constant innate opinions in the FJ model or evolving private opinions present in the EPO model. In general, both time-lag effects and hidden variables may operate simultaneously. This explains why delayed EPO models exhibit good predictive performance.

	\begin{table*}[tb]
        \centering
		\caption{
        Performance of the opinion dynamics models (best $\checkmark$, competitive $\star$).}
		\label{tab:summary}
		\begin{tabular}{llllllll}
			\toprule
			Model & Sum of & MAE & MAPE & RMSE & RMSE & RMSE & RMSE \\
			& residuals & & & Periods 1--10 & Period 11 & Period 12 & Periods 11--12 \\
			& & & & (in-sample) & (out-of-sample) & (out-of-sample) & (out-of-sample) \\
			\midrule
                FDG~model & 0.2000 & 0.1308 & 7.7957\% & 0.1461 & 0.1463 & 0.1927 & 0.1711 \\
                FJ~model & 0.1647 & 0.1257 & 7.5226\% & 0.1401 & 0.1330 $\star$ & 0.1704 & 0.1529 \\
                EPO~model & 0.0773 & 0.1129 & 6.7539\% & 0.1319 & 0.1422 & 0.1563 & 0.1494 \\
                Reduced EPO~model & 0.0879 & 0.1161 & 6.9463\% & 0.1347 & 0.1346 & 0.1457 $\star$ & \bf{0.1402 \checkmark} \\
                \midrule
                FDGM~model with lag~1 & 0.1754 & 0.1219 & 7.2959\% & 0.1423 & 0.1496 & 0.1512 & 0.1504 \\
                FDGM~model with lag~2 & 0.1412 & 0.0987 $\star$ & 5.8621\% $\star$ & 0.1242 $\star$ & 0.1238 $\star$ & 0.1641 & 0.1454 \\
                EPO~model with lag~1 & 0.0681 $\star$ & 0.1222 & 7.3173\% & 0.1411 & 0.1335 & 0.1559 & 0.1451 \\
                Reduced EPO~model with lag~1 & 0.0783 & 0.1277 & 7.6704\% & 0.1434 & 0.1347 & 0.1476 $\star$ & 0.1413 $\star$ \\
                EPO~model with lag~2 &  \bf{0.0530 \checkmark} & 0.0883 $\star$ & 5.2564\% $\star$ & 0.1154 $\star$ & 0.1584 & \bf{0.1340 \checkmark} & 0.1467 \\
                Reduced EPO~model with lag~2 & 0.0639 $\star$ & \bf{0.0854 \checkmark} & \bf{5.0497\% \checkmark} & \bf{0.1114 \checkmark} & \bf{0.1208 \checkmark} & 0.1641 & 0.1441 $\star$ \\
			\bottomrule
		\end{tabular}
	\end{table*}

\subsection{Fitting the Averaging-Based Models} 

We identify the unknown parameters as described in Section~\ref{subsec:ls}. Table~\ref{tab:summary} summarizes the models' in-sample goodness-of-fit and out-of-sample performance, measured by Mean Absolute Error (MAE), Mean Absolute Percentage Error (MAPE), and Root Mean Squared Error (RMSE). The out-of-sample evaluation uses periods 11 and 12.

\begin{figure}[h!]
        \small
		\centering
		\begin{tikzpicture}
			\begin{axis}
				[legend columns=4,
				legend style={at={(0.5,-0.25)},anchor=north},
				xlabel={Period, $t$},
				ylabel={Observed and predicted sentiments},
				xlabel near ticks,
				ylabel near ticks,
				xmin=1,
				xmax=12,
				ymin=0.38,
				ymax=0.85,
				xtick={1,2,3,4,5,6,7,8,9,10,11,12},
				ytick={0.4,0.5,0.6,0.7,0.8},
				width=8.5cm,
				height=6cm,
				legend cell align=left,
				legend style={draw=none, font=\scriptsize}
				]
				\pgfplotstableread[col sep=&, row sep=\\]{
					1 & 2 & 3 & 4 & 5 & 6 & 7 & 8 & 9 & 10 & 11 & 12\\
					0.542237 & 0.69269 & 0.719408 & 0.656396 & 0.662846 & 0.72706 & 0.582414 & 0.577169 & 0.658991 & 0.808328 & & \\
					0.690464 & 0.685991 & 0.626356 & 0.593513 & 0.809609 & 0.665289 & 0.529654 & 0.678874 & 0.659458 & 0.682581 & & \\
					0.640777 & 0.601049 & 0.617184 & 0.593332 & 0.643933 & 0.648383 & 0.601566 & 0.568535 & 0.610053 & 0.552651 & & \\
					0.687817 & 0.623997 & 0.529495 & 0.621614 & 0.649233 & 0.708136 & 0.598221 & 0.604527 & 0.607142 & 0.600858 & & \\
					0.473992 & 0.497273 & 0.513026 & 0.506612 & 0.546008 & 0.594167 & 0.523988 & 0.545223 & 0.571942 & 0.573838 & & \\
					0.740787 & 0.669839 & 0.666616 & 0.742431 & 0.571651 & 0.73716 & 0.655859 & 0.636605 & 0.669664 & 0.710904 & & \\
					0.551033 & 0.557094 & 0.537976 & 0.49017 & 0.632734 & 0.766128 & 0.574511 & 0.653399 & 0.694449 & 0.679584 & & \\
					& & & & & & & & & 0.808328 & 0.775289 & 0.713721 \\
					& & & & & & & & & 0.682581 & 0.709356 & 0.636683 \\
					& & & & & & & & & 0.552651 & 0.600009 & 0.499918 \\
					& & & & & & & & & 0.600858 & 0.592896 & 0.627276 \\
					& & & & & & & & & 0.573838 & 0.484829 & 0.561077 \\
					& & & & & & & & & 0.710904 & 0.679262 & 0.784448 \\
					& & & & & & & & & 0.679584 & 0.680478 & 0.708864 \\
					& & & & & & & & & 0.808328 & 0.692155 & 0.689703 \\
                    & & & & & & & & & 0.682581 & 0.69206 & 0.685445 \\
                    & & & & & & & & & 0.552651 & 0.577721 & 0.579021 \\
                    & & & & & & & & & 0.600858 & 0.574334 & 0.5952 \\
                    & & & & & & & & & 0.573838 & 0.584956 & 0.594069 \\
                    & & & & & & & & & 0.710904 & 0.684408 & 0.688682 \\
                    & & & & & & & & & 0.679584 & 0.677213 & 0.682164 \\
				}\datatable;
				\pgfplotstabletranspose\datatable{\datatable};
				\addplot[mark=o, blue!60!black, thick, opacity=0.2, forget plot] table [x index=1, y index=2]{\datatable};
				\addplot[mark=o, red!90!black, thick, opacity=0.2, forget plot] table [x index=1, y index=3]{\datatable};
				\addplot[mark=o, orange!80!yellow, thick, opacity=0.2, forget plot] table [x index=1, y index=4]{\datatable};
				\addplot[mark=o, green!60!black, thick, opacity=0.2, forget plot] table [x index=1, y index=5]{\datatable};
				\addplot[mark=o, magenta!90!orange, thick, opacity=0.2, forget plot] table [x index=1, y index=6]{\datatable};
				\addplot[mark=o, cyan!90!black, thick, opacity=0.2, forget plot] table [x index=1, y index=7]{\datatable};
				\addplot[mark=o, green!30!yellow, thick, opacity=0.2, forget plot] table [x index=1, y index=8]{\datatable};
				\addplot[mark=o, blue!60!black, thick] table [x index=1, y index=9]{\datatable};\addlegendentry{$x^e_1(t)\quad$}
				\addplot[mark=o, red!90!black, thick] table [x index=1, y index=10]{\datatable};\addlegendentry{$x^e_2(t)\quad$}
				\addplot[mark=o, orange!80!yellow, thick] table [x index=1, y index=11]{\datatable};\addlegendentry{$x^e_3(t)\quad$}
				\addplot[mark=o, green!60!black, thick] table [x index=1, y index=12]{\datatable};\addlegendentry{$x^e_4(t)\quad$}
				\addplot[mark=o, blue!60!black, thick, densely dashed] table [x index=1, y index=16]{\datatable};\addlegendentry{$\widehat{x}^e_1(t)\quad$}
				\addplot[mark=o, red!90!black, thick, densely dashed] table [x index=1, y index=17]{\datatable};\addlegendentry{$\widehat{x}^e_2(t)\quad$}
				\addplot[mark=o, orange!80!yellow, thick, densely dashed] table [x index=1, y index=18]{\datatable};\addlegendentry{$\widehat{x}^e_3(t)\quad$}
				\addplot[mark=o, green!60!black, thick, densely dashed] table [x index=1, y index=19]{\datatable};\addlegendentry{$\widehat{x}^e_4(t)\quad$}
				\addplot[mark=o, magenta!90!orange, thick] table [x index=1, y index=13]{\datatable};\addlegendentry{$x^e_5(t)\quad$}
				\addplot[mark=o, cyan!90!black, thick] table [x index=1, y index=14]{\datatable};\addlegendentry{$x^e_6(t)\quad$}
				\addplot[mark=o, green!30!yellow, thick] table [x index=1, y index=15]{\datatable};\addlegendentry{$x^e_7(t)\quad$}
				\addlegendimage{empty legend}\addlegendentry{}
				\addplot[mark=o, magenta!90!orange, thick, densely dashed] table [x index=1, y index=20]{\datatable};\addlegendentry{$\widehat{x}^e_5(t)\quad$}
				\addplot[mark=o, cyan!90!black, thick, densely dashed] table [x index=1, y index=21]{\datatable};\addlegendentry{$\widehat{x}^e_6(t)\quad$}
				\addplot[mark=o, green!30!yellow, thick, densely dashed] table [x index=1, y index=22]{\datatable};\addlegendentry{$\widehat{x}^e_7(t)\quad$}
			\end{axis}
		\end{tikzpicture}
		\caption{Reduced EPO Model: Observed and two-period ahead predicted sentiments, $x^e_b(t)=\sigma_b(t)$ and $\widehat{x}^e_b(t)$.}
		\label{fig:forecast}
	\end{figure}
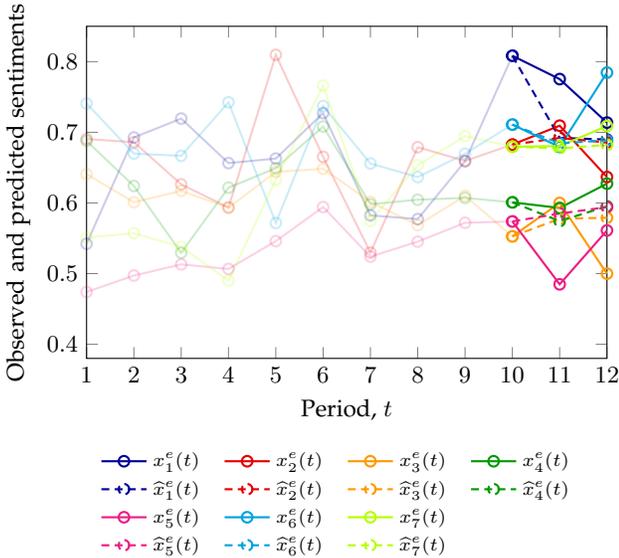

Anticipating the detailed comparison in the next subsection, we note some key findings. 
The delayed EPO models exhibit the best in-sample fit, which aligns with our finding from the range violation indices that audience sentiments have lagged effects extending beyond the immediate next period.
At the same time, introducing delays directly into the FDG model (as in the FDGM) does not significantly improve in-sample performance, suggesting that the presence of unobserved latent variables is more important than memory effects.
Interestingly, for out-of-sample prediction, the undelayed reduced EPO model achieves the lowest RMSE. Fig.~\ref{fig:forecast} illustrates the observed and two-period-ahead predicted sentiments using the reduced EPO~model, demonstrating the best predictive performance.

Further numerical details are presented in the exhibits below. The numerical values of estimated parameters are collected in Table~\ref{tab:models-fit}.
Fig.~\ref{fig:graphs} shows the heatmaps and network graphs for the fitted matrices $\widehat{W}$ and $\widehat{A}$.  Fig.~\ref{fig:fitted-sentiments-all} illustrates the dynamics of fitted sentiments for each model. The next subsection summarizes findings derived from these results.

\color{black}

\subsection{Observations and Discussion}

\paragraph*{\bf Sparsity} A first observation is that the fitted matrices $\widehat W$ (and $\widehat A$ in EPO models) exhibit emergent sparsity despite the absence of explicit sparsity constraints such as a pre-specified influence graph\footnote{The dependency structure among blog audiences is unknown a priori, and audience overlap cannot be determined since Weibo API discloses only a small fraction of the subscribers for each blog.} or sparsity-inducing penalty terms in the cost function. Entries below $< 10^{-5}$, reported as zeros, are more than 1000 times smaller than the minimal non-zero entries, indicating a clear separation between influential and negligible connections.

\begin{figure*}[h!]
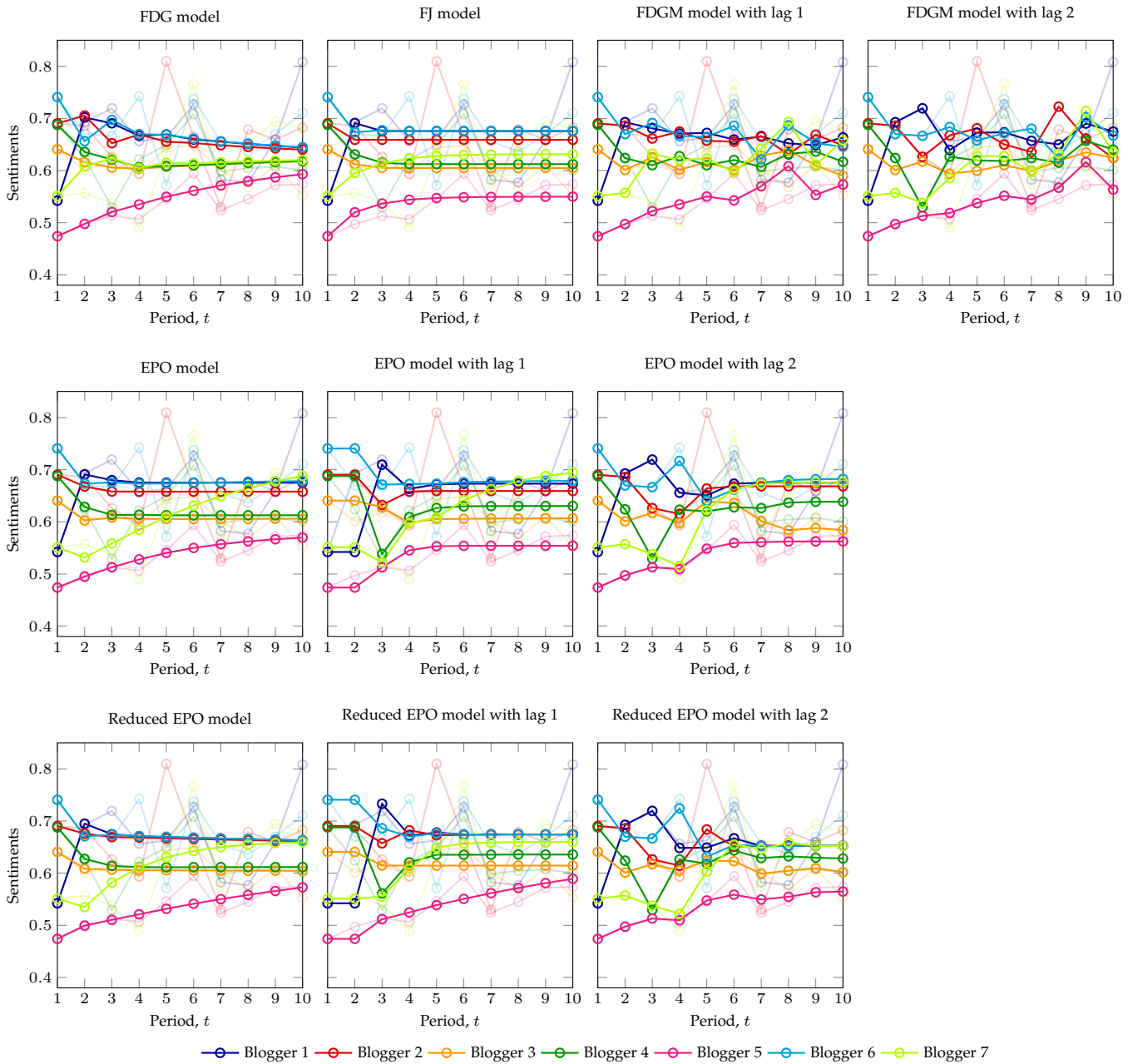

		\scriptsize
		\centering

		\caption{Sentiments for the parameters from Table~\ref{tab:models-fit}. Transparent line: measured sentiment $\sigma_b$; solid lines: model estimate $\widehat x_b^e$.}
		\label{fig:fitted-sentiments-all}
	\end{figure*}

\paragraph*{\bf Single-layer averaging}
The classical FDG model predictably performs worst, as it both exhibits rapid convergence to consensus (not observed in the sentiments) and requires that the opinion range not expand over time (a condition violated in our data; see Fig.~\ref{fig:mu}). The FJ model, which introduces constant ``innate'' opinions, slightly improves performance but still predicts rapid convergence to equilibria (Fig.~\ref{fig:fitted-sentiments-all}) -- a pattern not observed in the data. Meanwhile, the FDGM model with lag $2$ -- which uses delayed opinion vectors rather than static innate opinions -- demonstrates substantially better in-sample performance and ranks second for one-step-ahead prediction.
This suggests that single-layer averaging models, inheriting the structure of the FDG model, might be applicable for modeling sentiments; however, time lag effects need to be accommodated. These effects may differ across blogs: e.g., $\widehat{S}_{11}=1$ for the FDGM with 
$\tau=2$ indicates that sentiment updates in blog $1$ are not influenced by lagged opinions from previous steps.
Finally, we note that the FDGM represents the minimal time-lagged model. More generally, one could incorporate a sliding window of past opinions $x(t-1), x(t-2), \ldots, x(t-\tau)$ rather than a single lag, thereby introducing additional parameters.

\paragraph*{\bf Two-layer averaging} The EPO model, introducing ``hidden'' opinions for each macro-agent, demonstrates high consistency with the data, with performance improving further when time lags are incorporated. This may appear to be a simple consequence of increasing model complexity. However, many of the additional parameters have negligible influence. Beyond the emergent sparsity in $\widehat W$ and $\widehat A$,
many rows of $\widehat A$ and variables $\widehat x_b^e$ are ineffective since $\varphi_{bb}=1$ (which means that $\widehat x_b^e=\widehat x_b$).

\begin{figure*}[h!]
		\centering
		\includegraphics[width=0.95\textwidth]{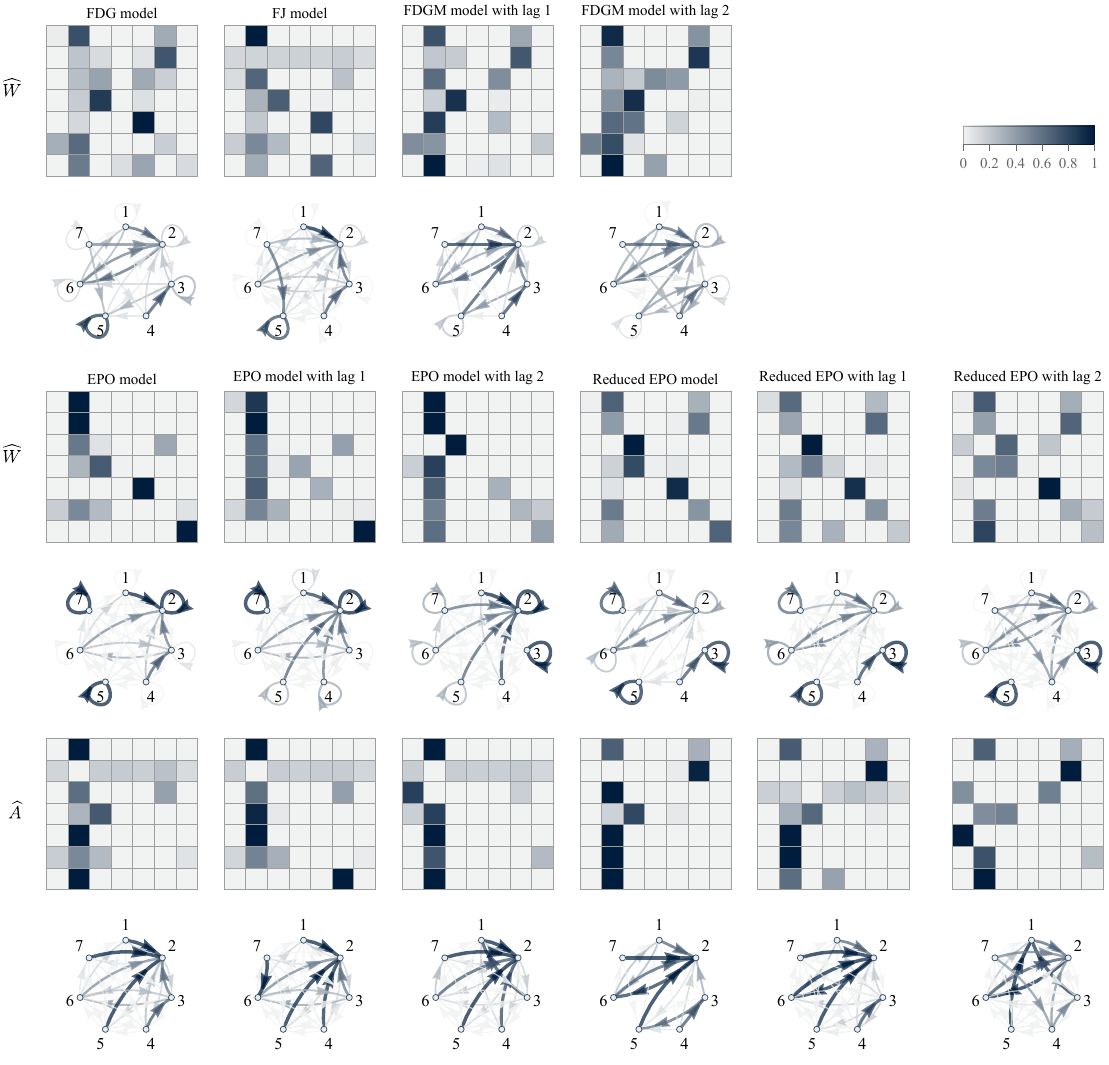}
		\caption{Fitted matrices $\widehat{W}$ and $\widehat{A}$ as heatmaps and their associated graphs.}
		\label{fig:graphs}
	\end{figure*}

Even for the most complex EPO models (in terms of parameters), good in-sample performance is not merely overfitting, as evidenced by sustained out-of-sample performance. For instance, the full EPO with lag $\tau=2$ delivers the best prediction for period $t=12$. However, reduced EPO models ($S=I$, $z$ ineffective) match or outperform full models despite having fewer parameters, with similar dependency structures (Fig.~\ref{fig:graphs}). Moreover, the delayed ($\tau=2$) reduced EPO achieves the best in-sample and one-step prediction, while the undelayed version excels at two-step-ahead forecasts. This suggests the FJ ``innate'' opinion mechanism is unnecessary; reduced EPO models, generalizing the FDG model, suffice to describe sentiments.

The heatmaps in Fig.~\ref{fig:graphs} show that the interblog influence graphs maintain similar structure across reduced EPO models with best predictive performance: Bloggers~3 and~5 heavily influence their commenters, with strong inter-content influence between Bloggers~2 and~6.

Interestingly, in many EPO models, the private opinion layer remains unused or weakly influences public opinions. We illustrate using the reduced EPO model, which achieves the best out-of-sample RMSE (0.1402 for periods 11--12). Except for Bloggers~3 and~5, users consistently express genuine opinions ($\widehat{\varphi}_b = 1$ for $b \neq 3, 5$); for these two bloggers, users provide authentic feedback in only $\sim$85\% of cases. Furthermore, feedback for Bloggers~3, 5, and to some extent~7 is primarily content-driven ($\widehat{w}_{33} = 1$, $\widehat{w}_{55} = 0.9271$, $\widehat{w}_{77} = 0.6664$), contrasting sharply with Bloggers~1 and~4, where sentiment is largely influenced by other bloggers ($\widehat{w}_{11} = 0.0160$, $\widehat{w}_{44} = 0.0206$). Similar patterns across other EPO models support the conclusion that users generally express genuine opinions when commenting.

	\begin{table*}[]
    \centering
    \caption{Fitted parameters estimated from periods 1--10. Entries are rounded to $10^{-5}$; zeros represent values $< 10^{-5}$.}
		\label{tab:models-fit}
		\small
    \begin{tabular}{>{\raggedright\arraybackslash}m{\columnwidth}|>{\raggedright\arraybackslash}m{\columnwidth}}
		\toprule
        FDG model

        \qquad$\begin{aligned}\widehat{W} =
			\begin{psmallmatrix}
				0.0244 & 0.6763 & 0 & 0 & 0 & 0.2993 & 0 \\
				0 & 0.1924 & 0.0989 & 0 & 0.0589 & 0.6498 & 0 \\
				0 & 0.2177 & 0.3311 & 0 & 0.3025 & 0.1487 & 0 \\
				0 & 0.1601 & 0.7622 & 0 & 0.0777 & 0 & 0 \\
				0 & 0.1097 & 0 & 0 & 0.8903 & 0 & 0 \\
				0.2813 & 0.5685 & 0 & 0 & 0 & 0.1502 & 0 \\
				0 & 0.4987 & 0 & 0.0847 & 0.3206 & 0 & 0.0960 \\
			\end{psmallmatrix}
            \phantom{\widehat{S}} & \phantom{\textrm{diag}\{\scalebox{0.8}{$0.4861,0,0.3347,0.5676,0.5509,0.6756,0.8905$}\}} \\
			\phantom{\widehat{z}} & \phantom{(\scalebox{0.8}{$0.6915,0.6590,0.5748,0.6026,0.5231,0.7394,1$})}
            \end{aligned}$
        &
        FJ model\vskip1mm

        \qquad $\begin{aligned}
				\widehat{W} &=
				\begin{psmallmatrix}
					0 & 1 & 0 & 0 & 0 & 0 & 0 \\
					0.1268 & 0.1422 & 0.1493 & 0.1554 & 0.1422 & 0.1756 & 0.1085 \\
					0.1051 & 0.6610 & 0 & 0.0029 & 0 & 0.2310 & 0 \\
					0 & 0.2920 & 0.6933 & 0 & 0.0147 & 0 & 0 \\
					0 & 0.2000 & 0 & 0 & 0.8000 & 0 & 0 \\
					0.1731 & 0.4940 & 0.2603 & 0 & 0 & 0 & 0.0726 \\
					0 & 0.3311 & 0 & 0 & 0.6689 & 0 & 0 \\
				\end{psmallmatrix}\\
				\widehat{S} &= \textrm{diag}\{\scalebox{0.8}{$0.4861,0,0.3347,0.5676,0.5509,0.6756,0.8905$}\} \\
				\widehat{z} &= (\scalebox{0.8}{$0.6915,0.6590,0.5748,0.6026,0.5231,0.7394,1$}) \\
			\end{aligned}$\\
\bottomrule
    \end{tabular}
\end{table*}

	\section{Conclusions and Future Works}\label{sec:conclusions}
	
	  This work contributes to sentiment dynamics modeling on social media by analyzing sentiment evolution on Weibo. A fine-tuned BERT model (\emph{bert-base-chinese}) captures and quantifies follower sentiments in response to influencer posts, offering insights into online community dynamics. We analyze several influential bloggers (each with over one million followers), tracking sentiment fluctuations toward topic-specific content over twelve 15-day periods. These sentiment values, interpreted as collective ``opinions'' of each blog's audience, exhibit clear \emph{averaging} dynamics: new opinions consistently fall within the range of earlier values (Fig.~\ref{fig:mu}), sometimes with 1--2 period lags. This delay likely reflects asynchronous user engagement or responses to earlier posts. Among several established opinion formation frameworks we evaluated, a modification of the EPO (expressed-private opinion) model best fit the data.
	

     The alignment between empirical data and averaging-based opinion dynamics models underscores their practical value for capturing sentiment evolution. These parsimonious models offer key advantages: they are interpretable, well-established, and enable direct learning of influence matrices from data using dynamical systems identification tools~\cite{ravazzi2021learning}. Once calibrated, they support sentiment trajectory prediction, identification of key influencers, and simulation of targeted interventions. While our study focuses on Weibo and Huawei-related content, the methodological framework -- sentiment extraction, user aggregation into macro-agents, and model calibration via opinion dynamics -- is topic-agnostic and applicable to other subjects and platforms. The two-level aggregation procedure could be refined -- for instance, by incorporating deeper comment levels. 

     \paragraph*{\bf Avenues for future research} First, more macro-agents could be introduced, such as smaller blogs or individual users. While the identification procedure extends in principle, this poses technical challenges. Expanding the analysis requires longer time frames due to Weibo's six-month data retention and lower commenting frequency in smaller blogs. Additionally, sentiment analysis noise becomes more pronounced with smaller follower bases. Individual sentiment estimation is particularly challenging given variability in writing styles and expressive behaviors; statistically meaningful insights emerge more reliably from large-group aggregation. Finally, least squares identification becomes computationally demanding as influence matrices grow. We leave these challenges for future work.
    
    Second, broader classes of dynamical models can be considered. While we use deterministic opinion dynamics with ordinary least squares (OLS) estimation, alternative approaches for multivariate time series analysis -- such as vector autoregression (VAR), maximum likelihood estimation (MLE), or Bayesian inference -- are viable. Our OLS approach effectively calibrates a low-order VAR model with additional constraints (e.g., stochastic matrices). Relaxing these constraints may improve predictive accuracy but sacrifices the guarantee that predictions remain in $[0,1]$, which is guaranteed by averaging-based models. Advanced statistical methods impose specific distributional assumptions, whereas our approach requires only observational data.

    While we prioritized compact models with minimal, easily identifiable parameters, one could also employ advanced opinion formation models with nonlinear dynamics (e.g., bounded confidence~\cite{Bernardo2024survey}, biased assimilation~\cite{Dandekar:2013}) or multiple delays (static or time-varying). However, such extensions require specialized identification techniques, presenting additional research challenges.

	\bibliographystyle{IEEEtran}
	\bibliography{bibfile,emotions}

@article{degroot1974reaching,
	title={Reaching a consensus},
	author={DeGroot, Morris H},
	journal={Journal of the American Statistical Association},
	volume={69},
	number={345},
	pages={118--121},
	year={1974},
	publisher={Taylor \& Francis Group}
}

@incollection{Abelson:1964,
    author={R.P. Abelson},
    title={Mathematical   Models  of  the  Distribution  of  Attitudes under Controversy},
    booktitle={Contributions  to Mathematical Psychology},
    editor={N. Frederiksen and H. Gulliksen},
    year={1964},
    pages={142--160},
    publisher={Holt, Rinehart \& Winston, Inc},
    address={New York},
}

@article{french1956formal,
	title={A formal theory of social power.},
	author={French Jr, John RP},
	journal={Psychological review},
	volume={63},
	number={3},
	pages={181},
	year={1956},
	publisher={American Psychological Association}
}

@article{FJ99,
  author = {N.~E.~Friedkin and E.~C.~Johnsen},
  journal = {Adv. Group Processes},
  pages = {1--29},
  title = {{Social influence networks and opinion
change}},
  volume = 16,
  year = 1999
}

@article{nematollahzadeh2020learning,
	title={Learning influential cognitive links in social networks by a new hybrid model for opinion dynamics},
	author={Nematollahzadeh, Seyed Mahmood and Ozgoli, Sadjaad and Haghighi, Mohammad Sayad and Jolfaei, Alireza},
	journal={IEEE Transactions on Computational Social Systems},
	volume={8},
	number={5},
	pages={1262--1271},
	year={2020},
	publisher={IEEE}
}

@article{Dandekar:2013,
    author = {Pranav Dandekar  and Ashish Goel  and David T. Lee },
    title = {Biased assimilation, homophily, and the dynamics of polarization},
    journal = {Proceedings of the National Academy of Sciences},
    volume = {110},
    number = {15},
    pages = {5791-5796},
    year = {2013},
}

@article{Friedkin:2015,
    author = {N. Friedkin},
    title  = {The Problem of Social Control and Coordination of Complex Systems in Sociology: A Look at the Community Cleavage Problem},
    journal={IEEE Control Syst. Mag.},
    volume={35},
    number={3},
	pages={40--51},
    year={2015}
}

@article{proskurnikov2018tutorial,
  title={A tutorial on modeling and analysis of dynamic social networks. {P}art {II}},
  author={A.~Proskurnikov and R.~Tempo},
  journal={Annu. Rev. Control},
  volume={45},
  pages={166--190},
  year={2018},
  publisher={Elsevier}
}

@article{proskurnikov2017tutorial,
	title={A tutorial on modeling and analysis of dynamic social networks. {P}art {I}},
	 author={A.~Proskurnikov and R.~Tempo},
	journal={Annu. Rev. Control},
	volume={43},
	pages={65--79},
	year={2017}
}

@article{ye2019influence,
	title={An influence network model to study discrepancies in expressed and private opinions},
	author={Ye, Mengbin and Qin, Yuzhen and Govaert, Alain and Anderson, Brian DO and Cao, Ming},
	journal={Automatica},
	volume={107},
	pages={371--381},
	year={2019},
	publisher={Elsevier}
}

@article{FriedProsk2019,
    author={Friedkin, N.E. and Proskurnikov, A.V. and Mei, W. and Bullo, F.},
    title={Mathematical Structures in Group Decision-Making on Resource Allocation Distributions},
    journal={Scientific Reports},
    volume={5},
    number={1},
    pages={1377},
    year={2019},
}

@article{mastroeni2019agent,
  title={Agent-based models for opinion formation: A bibliographic survey},
  author={Mastroeni, Loretta and Vellucci, Pierluigi and Naldi, Maurizio},
  journal={IEEE Access},
  volume={7},
  pages={58836--58848},
  year={2019},
  publisher={IEEE}
}

@article{mosteller1964inference,
  title={Inference and disputed authorship: the Federalist},
  author={Mosteller, Frederick and Wallace, David Lee},
  journal={(No Title)},
  year={1964}
}

@article{Noorazar2020,
  author = {Hossein Noorazar},
  title = {Recent Advances in Opinion Propagation Dynamics: A 2020 Survey},
  journal = {The European Physical Journal Plus},
  year = {2020},
  volume={135},
  pages={521},
}

@article{FRIEDKINPROBULLO:2021,
    title = {Group dynamics on multidimensional object threat appraisals},
    journal = {Social Networks},
    volume = {65},
    pages = {157--167},
    year = {2021},
    author = {Noah E. Friedkin and Anton  Proskurnikov and Francesco Bullo},
}

@article{ravazzi2021learning,
  title={Learning Hidden Influences in Large-Scale Dynamical Social Networks: A Data-Driven Sparsity-Based Approach, in Memory of {R}oberto {T}empo},
  author={Ravazzi, Chiara and Dabbene, Fabrizio and Lagoa, Constantino and Proskurnikov, Anton V},
  journal={IEEE Control Syst. Mag.},
  volume={41},
  number={5},
  pages={61--103},
  year={2021},
  publisher={IEEE}
}

@ARTICLE{Wai2016,
  author={Wai, Hoi-To and Scaglione, Anna and Leshem, Amir},
  journal={IEEE Transactions on Signal and Information Processing over Networks}, 
  title={Active Sensing of Social Networks}, 
  year={2016},
  volume={2},
  number={3},
  pages={406-419},
}

@article{hochreiter1997long,
  title={Long Short-term Memory},
  author={Hochreiter, S},
  journal={Neural Computation MIT-Press},
  year={1997}
}

@article{schuster1997bidirectional,
  title={Bidirectional recurrent neural networks},
  author={Schuster, Mike and Paliwal, Kuldip K},
  journal={IEEE Transactions on Signal Processing},
  volume={45},
  number={11},
  pages={2673--2681},
  year={1997},
  publisher={IEEE}
}

@inproceedings{devlin2018bert,
  author       = {Jacob Devlin and
                  Ming{-}Wei Chang and
                  Kenton Lee and
                  Kristina Toutanova},
  editor       = {Jill Burstein and
                  Christy Doran and
                  Thamar Solorio},
  title        = {{BERT:} Pre-training of Deep Bidirectional Transformers for Language
                  Understanding},
  booktitle    = {Proceedings of the 2019 Conference of the North American Chapter of
                  the Association for Computational Linguistics: Human Language Technologies,
                  {NAACL-HLT} 2019},
  pages        = {4171--4186},  
  year         = {2019},  
  doi          = {10.18653/V1/N19-1423},  
}

@data{weibosenti,
	doi = {10.21227/abj8-y636},
	url = {https://dx.doi.org/10.21227/abj8-y636},
	author = {Sun, Maosong},
	publisher = {IEEE Dataport},
	title = {Weibo\_senti\_100k and {THUC}News},
	year = {2022} 
}

@book{deonna2008,
  title     = "The Emotions: A Philosophical Introduction",
  author    = "Deonna, Julien and Teroni, Fabrice",
  year      = 2008,
  publisher = "Routledge",
  address   = "London"
}

@article{wong2016quantifying,
  title={Quantifying political leaning from tweets, retweets, and retweeters},
  author={Wong, Felix Ming Fai and Tan, Chee Wei and Sen, Soumya and Chiang, Mung},
  journal={IEEE Transactions on Knowledge and Data Engineering},
  volume={28},
  number={8},
  pages={2158--2172},
  year={2016},
  publisher={IEEE}
}

@article{zhao2024opinion,
  author    = {Chi Zhao and Elena Parilina},
  title     = {Opinion Dynamics in Two-Layer Networks with Hypocrisy},
  journal   = {Journal of the Operations Research Society of China},
  year      = {2024},
  volume    = {12},
  number    = {1},
  pages     = {109--132},
}

@book{hatfield1993emotional,
  title={Emotional Contagion},
  author={Hatfield, Elaine and Cacioppo, John T. and Rapson, Richard L.},
  year={1993},
  publisher={Cambridge University Press}
}

@article{mehrabian1996pad,
  author    = {Albert Mehrabian},
  title     = {Pleasure-arousal-dominance: A general framework for describing and measuring individual differences in temperament},
  journal   = {Current Psychology},
  volume    = {14},
  number    = {4},
  pages     = {261--292},
  year      = {1996},
  doi       = {10.1007/BF02686918},
}

@article{barsade2002ripple,
  title={The Ripple Effect: Emotional Contagion in Groups},
  author={Barsade, Sigal G.},
  journal={Administrative Science Quarterly},
  year={2002},
  volume={47},
  number={4},
  pages={644-675},
  publisher={SAGE Publications}
}

@article{russell2003core,
  author    = {Russell, James A.},
  title     = {Core affect and the psychological construction of emotion},
  journal   = {Psychological Review},
  volume    = {110},
  number    = {1},
  pages     = {145--172},
  year      = {2003},
  doi       = {10.1037/0033-295X.110.1.145},
}

@article{Rey2010,
  author = {Jos{\'e}-Manuel Rey},
  title = {A Mathematical Model of Sentimental Dynamics Accounting for Marital Dissolution},
  journal = {PLoS ONE},
  volume = {5},
  number = {3},
  pages = {e9881},
  year = {2010},
}

@inproceedings{Macropol2013,
  author = {Kathy Macropol and Petko Bogdanov and Ambuj K. Singh and Linda R. Petzold and Xifeng Yan},
  title = {I act, therefore {I} judge: Network sentiment dynamics based on user activity change},
  booktitle = {Proceedings of the 2013 IEEE/ACM International Conference on Advances in Social Networks Analysis and Mining (ASONAM)},
  pages = {396--402},
  year = {2013},
}

@article{he2014djst,
    author = {He, Yulan and Lin, Chenghua and Gao, Wei and Wong, Kam-Fai},
    title = {Dynamic joint sentiment-topic model},
    year = {2014},    
    volume = {5},
    number = {1},
    journal = {ACM Trans. Intell. Syst. Technol.},    
    articleno = {6},
    numpages = {21},
}

@article{kramer2014emotional,
  title={Experimental Evidence of Massive-Scale Emotional Contagion through Social Networks},
  author={Kramer, Adam D. I. and Guillory, Jamie E. and Hancock, Jeffrey T.},
  journal={Proceedings of the National Academy of Sciences},
  year={2014},
  volume={111},
  number={24},
  pages={8788-8790},
}

@article{franke2014,
title = {Aggregate sentiment dynamics: A canonical modelling approach and its pleasant nonlinearities},
journal = {Structural Change and Economic Dynamics},
volume = {31},
pages = {64-72},
year = {2014},
author = {Reiner Franke},
}

@article{Bakshy2015,
author = {Eytan Bakshy  and Solomon Messing  and Lada A. Adamic},
title = {Exposure to ideologically diverse news and opinion on {F}acebook},
journal = {Science},
volume = {348},
number = {6239},
pages = {1130-1132},
year = {2015},
}

@book{Liu2015,
  author = {Bing Liu},
  title = {Sentiment Analysis: Mining Opinions, Sentiments, and Emotions},
  publisher = {Cambridge University Press},
  year = {2015},
  address = {Cambridge, UK},
}

@article{Barsade2018,
title = {Emotional contagion in organizational life},
journal = {Research in Organizational Behavior},
volume = {38},
pages = {137-151},
year = {2018},
author = {Sigal G. Barsade and Constantinos G.V. Coutifaris and Julianna Pillemer},
}

@article{rosenbusch2019multilevel,
  title={Multilevel Emotion Transfer on {Y}ou{T}ube: Disentangling the Effects of Emotional Contagion and Homophily on Video Audiences},
  author={Rosenbusch, Hannes and Evans, Anthony M. and Zeelenberg, Marcel},
  journal={Social Psychological and Personality Science},
  year={2019},
  volume={10},
  number={8},
  pages={1028-1035},
}

@ARTICLE{das2019,
  author={Das, Rajkumar and Kamruzzaman, Joarder and Karmakar, Gour},
  journal={IEEE Transactions on Computational Social Systems}, 
  title={Opinion Formation in Online Social Networks: Exploiting Predisposition, Interaction, and Credibility}, 
  year={2019},
  volume={6},
  number={3},
  pages={554-566},
}

@article{anderson2020ifac,
    title = {Dynamical Networks of Social Influence: Modern Trends and Perspectives},
    journal = {IFAC-PapersOnLine},
    volume = {53},
    number = {2},
    pages = {17616-17627},
    year = {2020},
    author = {Brian D.O. Anderson and Fabrizio Dabbene and Anton V. Proskurnikov and Chiara Ravazzi and Mengbin Ye},
}

@article{herrando2021emotional,
  title={Emotional Contagion: A Brief Overview and Future Directions},
  author={Herrando, Carolina and Constantinides, Efthymios},
  journal={Frontiers in Psychology},
  volume={12},
  pages={712606},
  year={2021},
  publisher={Frontiers},
}

@article{Naskar2020,
    author = {Naskar, Debashis and Singh, Sanasam Ranbir and Kumar, Durgesh and Nandi, Sukumar and Rivaherrera, Eva Onaindia de la},
    title = {Emotion Dynamics of Public Opinions on {T}witter},
    year = {2020},    
    volume = {38},
    number = {2},
    doi = {10.1145/3379340},
    journal = {ACM Trans. Inf. Syst.},
    month = mar,
    articleno = {18},
    numpages = {24},    
}

@article{zhang2021multilevel,
  title={Exploring Coevolution of Emotional Contagion and Behavior for Microblog Sentiment Analysis: A Deep Learning Architecture},
  author={Zhang, Qi and Zhang, Zufan and Yang, Maobin and Zhu, Lianxiang and Yang, Luxing},
  journal={Complexity},
  year={2021},
  volume={2021},
  pages={6630811},
}

@article{crocamo2021surveilling,
  title={Surveilling {COVID-19} Emotional Contagion on {T}witter by Sentiment Analysis},
  author={Crocamo, Cristina and Viviani, Marco and Famiglini, Lorenzo and Bartoli, Francesco and Pasi, Gabriella and Carr{\'a}, Giuseppe},
  journal={European Psychiatry},
  volume={64},
  number={1},
  year={2021},
}

@article{yin2021,
title = {Modelling the dynamic emotional information propagation and guiding the public sentiment in the {C}hinese {S}ina-microblog},
journal = {Applied Mathematics and Computation},
volume = {396},
pages = {125884},
year = {2021},
author = {Fulian Yin and Xinyu Xia and Xiaojian Zhang and Mingjia Zhang and Jiahui Lv and Jianhong Wu},
}

@article{Ahmed2021,
  author = {Md Shoaib Ahmed and Tanjim Taharat Aurpa and Md Musfique Anwar},
  title = {Detecting sentiment dynamics and clusters of {T}witter users for trending topics in {COVID-19} pandemic},
  journal = {PLoS ONE},
  volume = {16},
  number = {8},
  pages = {e0253300},
  year = {2021},
}

@article{nandwani2021review,
  author    = {Nandwani, Piyush and Verma, Rohit},
  title     = {A review on sentiment analysis and emotion detection from text},
  journal   = {Social Network Analysis and Mining},
  volume    = {11},
  number    = {1},
  pages     = {81},
  year      = {2021},
}

@article{lian2022,
    title = {An opinion dynamics model for unrelated discrete opinions},
    journal = {Knowledge-Based Systems},
    volume = {251},
    pages = {109133},
    year = {2022},
    author = {Ying Lian and Xuefan Dong},
}

@article{kozitsin2022,
	author = {Ivan V. Kozitsin},
	journal = {J. Math. Sociol.},
	number = {2},
	pages = {120--147},
	title = {Formal models of opinion formation and their application to real data: {E}vidence from online social networks},
	volume = {46},
	year = {2022}
}

@article{yin2022,
title = {Sentiment mutation and negative emotion contagion dynamics in social media: A case study on the {C}hinese {S}ina Microblog},
journal = {Information Sciences},
volume = {594},
pages = {118-135},
year = {2022},
author = {Fulian Yin and Xinyu Xia and Yanyan Pan and Yuwei She and Xiaoli Feng and Jianhong Wu},
}

@article{liu2022analysis,
  title={The analysis of dynamic emotional contagion in online brand community},
  author={Liu, D. and Zhang, S. and Li, Q.},
  journal={Frontiers in Psychology},
  volume={13},
  pages={946666},
  year={2022},
}

@article{patil2022,
  author    = {Rupali S. Patil and Satish R. Kolhe},
  title     = {Supervised classifiers with {TF-IDF} features for sentiment analysis of {M}arathi tweets},
  journal   = {Social Network Analysis and Mining},
  volume    = {12},
  number    = {1},
  pages     = {51},
  year      = {2022}
}

@article{VendrellFerran2022,
	author = {Vendrell Ferran,\'{I}ngrid},	
	journal = {Phenomenology and Mind},
	pages = {20--34},
	title = {Emotions and Sentiments: Two Distinct Forms of Affective Intentionality},
	volume = {23},
	year = {2022}
}

@ARTICLE{Zhang2023survey,
  author={Zhang, Wenxuan and Li, Xin and Deng, Yang and Bing, Lidong and Lam, Wai},
  journal={IEEE Transactions on Knowledge and Data Engineering}, 
  title={A Survey on Aspect-Based Sentiment Analysis: Tasks, Methods, and Challenges}, 
  year={2023},
  volume={35},
  number={11},
  pages={11019-11038},
}

@article{Meyer2023,
title = {High on Bitcoin: Evidence of emotional contagion in the {Y}ou{T}ube crypto influencer space},
journal = {Journal of Business Research},
volume = {164},
pages = {113850},
year = {2023},
issn = {0148-2963},
author = {Eva Andrea Meyer and Philipp Sandner and Bernard Cloutier and Isabell M. Welpe},
}

@ARTICLE{He2023,
  author={He, Qiang and Fang, Hui and Zhang, Jie and Wang, Xingwei},
  journal={IEEE Transactions on Knowledge and Data Engineering}, 
  title={Dynamic Opinion Maximization in Social Networks}, 
  year={2023},
  volume={35},
  number={1},
  pages={350-361},
}

@article{Grabisch2023,
author = {Grabisch, Michel and Mandel, Antoine and Rusinowska, Agnieszka},
title = {On the Design of Public Debate in Social Networks},
journal = {Operations Research},
volume = {71},
number = {2},
pages = {626-648},
year = {2023},
}

@article{vanHaeringen2023emotion,
  title     = {Emotion contagion in agent-based simulations of crowds: a systematic review},
  author    = {van Haeringen, E. S. and Gerritsen, C. and Hindriks, K. V.},
  journal   = {Autonomous Agents and Multi-Agent Systems},
  volume    = {37},
  number    = {6},
  year      = {2023},
}

@ARTICLE{vanHaeringen2024,
  author={van Haeringen, Erik Stefan and Veltmeijer, Emmeke Anna and Gerritsen, Charlotte},
  journal={IEEE Transactions on Affective Computing}, 
  title={Empirical Validation of an Agent-Based Model of Emotion Contagion}, 
  year={2024},
  volume={15},
  number={1},
  pages={273-284},
}

@article{Rimpy2024,
  author = {Rimpy, Dahiya and Dhankhar, Amita and Dhankhar, Akshika},
  title = {Sentimental Analysis of Social Networks: A Comprehensive Review (2018-2023)},
  journal = {Multidisciplinary Reviews},
  volume = {7},
  number = {7},
  pages = {2024126},
  year = {2024}
}

@article{Song2024,
  author    = {Xiaolei Song and Siliang Guo and Yichang Gao},
  title     = {Personality traits and their influence on Echo chamber formation in social media: a comparative study of {T}witter and {W}eibo},
  journal   = {Frontiers in Psychology},
  volume    = {15},
  pages     = {1323117},
  year      = {2024},
}

@ARTICLE{Zhou2024fj,
  author={Zhou, Xiaotian and Sun, Haoxin and Xu, Wanyue and Li, Wei and Zhang, Zhongzhi},
  journal={IEEE Transactions on Knowledge and Data Engineering}, 
  title={Friedkin-{J}ohnsen Model for Opinion Dynamics on Signed Graphs}, 
  year={2024},
  volume={36},
  number={12},
  pages={8313-8327},
}

@article{Bernardo2024survey,
title = {Bounded confidence opinion dynamics: A survey},
journal = {Automatica},
volume = {159},
pages = {111302},
year = {2024},
author = {Carmela Bernardo and Claudio Altafini and Anton Proskurnikov and Francesco Vasca},
}

@article{chen2025s3eir,
  title     = {Research on propagation dynamics of emotional contagion using {S3EIR} model based on multiple social media platforms},
  author    = {Chen, Hejie and Liu, Mingyu and Zhang, Yiqing and Wang, Hui},
  journal   = {Frontiers in Communication},
  year      = {2025},
  volume    = {10},
  pages     = {1582974},
}

@article{Xu2025_survey,
  title     = {Sentiment Diffusion in Online Social Networks: A Survey from the Computational Perspective},
  author    = {Xu, Han and Xu, Minghua and Deng, Xianjun and Wang, Bang},
  journal   = {ACM Computing Surveys},
  year      = {2025}, 
}

@article{Ilyas2025,
    title = {A systematic review of social media-based sentiment analysis in disaster risk management},
    journal = {International Journal of Disaster Risk Reduction},
    volume = {123},
    pages = {105487},
    year = {2025},
    author = {Bilal Ilyas and Ayyoob Sharifi},
}

	\appendix
    \begin{table*}[h]
        \centering
		\caption{Sentiment analysis algorithms: performance}
		\label{tab:comparison}
		\begin{tabular}{lllllll}
			\toprule
			Method & Dataset & Precision & Recall & F1 & Accuracy & AUC \\
			\midrule
			Naive Bayes & \emph{weibo\_senti\_100k} & 0.7444 & 0.8609 & 0.7984 & 0.7 & 0.6044 \\
			BiLSTM & \emph{weibo\_senti\_100k} & \textbf{0.8950 \checkmark} & 0.6174 & 0.7307 & 0.686 & 0.8071 \\  
			BERT & \emph{ChnSentiCorp-Htl-ba-10000} & 0.8754 & 0.8145 & 0.8438 & 0.792 & 0.8672 \\		BERT & \emph{weibo\_senti\_100k} & 0.8531 & \textbf{0.9594 \checkmark} & \textbf{0.9031 \checkmark} & \textbf{0.858 \checkmark} & \textbf{0.9469 \checkmark} \\
			\bottomrule
		\end{tabular}
	\end{table*}

	\begin{table*}[h]
        \centering
		\caption{The \emph{weibo\_senti\_100k} dataset}
		\label{tab:weibosenti100}
		\begin{CJK}{UTF8}{gbsn}
			\begin{tabular}{ll}
				\toprule
				Label & Review \\
				\midrule
				\ldots & \ldots \\
				1 & [给力]感谢所有支持雯婕的芝麻！[爱你]\\
				1 & 2013最后一天，在新加坡开心度过，向所有的朋友们问声：新年快乐！2014年，我们会更好[调...\\
				0 & 大中午出门办事找错路，曝晒中。要多杯具有多杯具。[泪][泪][汗]\\
				0 & 马航还会否认吗？到底在隐瞒啥呢？[抓狂]//@头条新闻: 转发微博\\
				0 & 克罗地亚球迷很爱放烟火！球又没进，就硝烟四起。[晕]\\
				1 & [抱抱]福芦 TangRoulou 吉祥书 8.8折优惠\texttt{>>>}http://t.cn/z...\\
				1 & 回复@钱旭明QXM:[嘻嘻][嘻嘻] //@钱旭明QXM:杨大哥[good][good][g...\\
				1 & 人家这脸长的!!!!!![哈哈]\\
				1 & 这个价不算高，和一天内训相比相差无几。。[哈哈]//@博通传媒v: 6个月！一个月工资1万，...\\
				0 & 终于收工啦，脚丫子快冻掉了[泪][泪]\\
				\ldots & \ldots \\
				\bottomrule
			\end{tabular}
		\end{CJK}
	\end{table*}

    
    To compare the models' performance on Chinese social media text, we use a manually labeled 2018 Weibo test dataset.\footnote{\url{https://github.com/dengxiuqi/weibo2018}.}
    The algorithm performance is evaluated using standard metrics such as accuracy, precision, recall, F1 score~\cite{patil2022}, and AUC. Results are summarized in Table~\ref{tab:comparison}.    

    \textbf{The Naive Bayes Model:} The implementation of this model is based on the \emph{SnowNLP}\footnote{\url{https://github.com/isnowfy/snownlp}.} library, specially designed for processing of simplified Chinese text.    

\begin{figure}[h]
		\centering
		\includegraphics[width=0.75\linewidth]{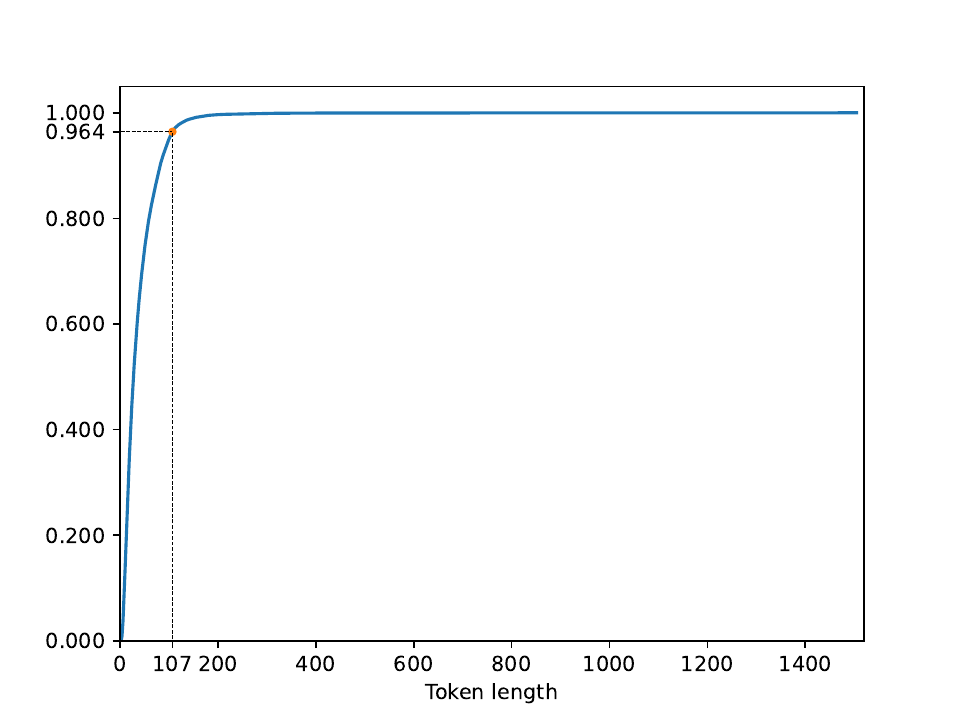}
		\caption{The CDF of a token length.}
		\label{fig:bilstm}
	\end{figure}

    \textbf{The BiLSTM Model:} We utilize the \emph{Keras}\footnote{\url{https://github.com/fchollet/keras}.} library to implement BiLSTM-based model the objective of enhancing the accuracy and efficiency of text processing. 
    The first layer of the model is the embedding layer, which transforms vocabulary indices into word vector matrices, a prerequisite for utilizing neural networks to process textual data. Based on the distribution of token length, as illustrated in Fig.~\ref{fig:bilstm}, the model is designed with an input sequence length of 107, with each word represented as a 300-dimensional vector. 
    
    The bidirectional layer allows the model to propagate information in both directions of the input sequence, thereby improving its information extraction capability. The long short-term memory (LSTM) layer serves as the core of the model, while the dense layer maps lower-dimensional representations to higher dimensions, ultimately producing a one-dimensional output suitable for binary classification. The structure of the model is summarized in Table~\ref{tab:bilstm-srtucture}.
    \begin{table}[h]
        \centering
		\caption{The structure of the BiLSTM model}\label{tab:bilstm-srtucture}		
		\begin{tabular}{lll}
			\toprule
            Layer Type & Output Shape & Param \#   \\
			\midrule
            Embedding & (None, 107, 300) & 58560600 \\
            Bidirectional & (None, 107, 64) & 85248 \\
            LSTM & (None, 16) & 5184 \\
            Dense & (None, 1) & 17 \\
			\bottomrule
		\end{tabular}
    \end{table}

	\textbf{The BERT Model:} The model is implemented using the Transformers library.\footnote{\url{https://github.com/huggingface/transformers}.}  We start from the pre-trained BERT model, namely \emph{bert-base-chinese} \cite{devlin2018bert}, which has been specifically developed for processing Chinese language texts. We fine-tune
    the pre-trained model using two sentiment annotation datasets: \emph{ChnSentiCorp-Htl-ba-10000} (Dataset~1) and \emph{weibo\_senti\_100k} (Dataset~2)
    focusing on hotel reviews and short Weibo texts, respectively. The \emph{weibo\_senti\_100k} dataset~\cite{weibosenti}, illustrated by Table~\ref{tab:weibosenti100}, contains over 100,000 anonymous pieces of Weibo text data with sentiment annotations. Approximately half of the texts are classified as positive (label~1), and the other half as negative (label~0). 
        
    The fine-tuning process involves adjusting the pre-trained \emph{bert-base-chinese} model to more accurately interpret the sentiment-specific characteristics and patterns present in the datasets. 
    The objective is to create a model that not only retains the robust language processing abilities of BERT but also excels in identifying and categorizing sentiments expressed in Chinese texts.
		




    \balance

\end{document}